# Equations of motion for the mass centers in a scalar theory of gravitation

Mayeul ARMINJON

*Laboratoire "Sols, Solides, Structures", Institut de Mécanique de Grenoble, B.P. 53, 38041 Grenoble cedex 9, France. E-mail: arminjon@hmg.inpg.fr*

A scalar theory of gravitation with a preferred reference frame (PRF) is considered, that accounts for special relativity and reduces to it if the gravitational field cancels. The gravitating system consists of a finite number of perfect-fluid bodies. An « asymptotic » post-Newtonian (PN) approximation scheme is used, allowing an explicit weak-field limit with all fields expanded. Exact mass centers are defined and their exact equations of motion are derived. The PN expansion of these equations is obtained: the zero-order equations are those of Newtonian gravity (NG), and the equations for the first-order (PN) corrections depend linearly on the PN fields. For PN corrections to the motion of the mass centers, especially in the solar system, one may assume « very-well-separated » rigidly moving bodies with spherical self-fields of the zero-order approximation. The PN corrections reduce then to a time integration and include spin effects, which might be significant. It is shown that the Newtonian masses are not correct zero-order masses for the PN calculations. An algorithm is proposed, in order to minimize the residual and to assess the velocity in the PRF.

## 1. Introduction

In 1905, Einstein suggested that the very notion of an absolute motion does not make sense in the physical world, and abolished the ether as « superfluous. » However, the « principle of relativity » was formulated in 1904 under this same name by Poincaré [1], in the following form: « the laws of physical phenomenons must be the same, whether for a fixed observer, as also for one dragged in a motion of uniform translation, so that we do not and cannot have any mean to discern whether or not we are dragged in a such motion. » Thus, for Poincaré, the notion of an absolute motion does make sense, even though « it seems that this impossibility to demonstrate the absolute motion be a general law in Nature » [2]. Moreover, Poincaré, as well as Lorentz, always reserved the possibility that the principle of relativity might be falsified by an observation, and indeed both considered an ether until the end of their lives (see *e.g.* Ref. 3).

The principle of relativity, as formulated either by Poincaré or by Einstein, leads to the same physical theory of *special* relativity (SR), which is one of the best-verified theories of physics. But SR does not involve gravitation. Although Einstein's theory of gravitation, which is the currently accepted theory of gravitation (see *e.g.* Will [4]), is named « general relativity » (GR), it is well-known since Fock [5] that the principle of relativity does not hold true in GR. For instance, it is clear that an observer *can* discern in a number of ways whether or not she is in a motion with respect to the static reference frame of Schwarzschild's space-time. It is true that, due to its manifest covariance under arbitrary coordinate changes, GR does not have any *a priori* preferred reference frame, in the following sense: if, in GR, we model the solar system (say) as *isolated*, we do not have to worry about the motion of the solar system with respect to whatever reference frame, and thus may choose our coordinate system such that, for instance, the Sun is at rest in that coordinate system. (In reality, however, the solar system is not isolated: it should be embedded in a large-scale model, thus the metric in the solar system should be « matched » with, say, the Robertson-Walker space-time metric representing the assumed macroscopically-homogeneous Universe.) But the point is that the experimental support of SR does not, as such, provide much constraint on the theory of gravitation: one may define a very large class of « relativistic » modifications of Newton's gravity (NG), by the condition that any of them should *reduce to SR when the gravitational field cancels.* (In GR, following Synge [6], this may be defined as the case where the Riemann tensor cancels, in which case the Minkowski space-time is indeed obtained.) If one defines in this way the « relativistic » character of a theory of gravitation, nothing forbids that a such theory may have an *a priori* privileged reference frame − thereby violating, not only the explicit form of the principle of relativity (as is already the case for GR), but its very spirit. If a privileged reference frame would manifest itself in the presence of a gravitational field (while remaining « hidden » in its absence), it would be easier to extend quantum mechanics to the situation with gravitation [7].

Many consider that, among relativistic theories of gravitation, it is unlikely that any preferred-frame theory could make sensible predictions. However, it has recently been proposed [8-10] a theory which, despite the simplicity which is conveyed to it by the fact that it is a *scalar* theory, and despite its preferred-frame character, has the following encouraging four properties: (**i**) it is relativistic, *i.e.,* it does reduce to SR when the gravitational field cancels. (**ii**) It has the correct Newtonian limit [11]. More precisely, one may define, within that theory, a small parameter $\varepsilon$ that characterizes the weakness of the gravitational field in the considered system and that is approximately equal to the ratio of the maximum orbital velocity in the system, to the velocity of light. And one may build an asymptotic scheme in which, up to the order $\varepsilon^2$ not included, the expanded equations reduce to NG [12-13]. This means that this theory is (at least) as good as NG in the solar system. (**iii**) It *predicts* [15] the observed acceleration of the cosmic expansion [16, 17] (**iv**) It recovers exactly the standard post-Newtonian (PN) predictions of GR as to the gravitational effects on light rays [14]. This is a crucial point, because those effects represent the best-verified consequences of GR, and nowadays the most important ones practically. Yet it is not enough that this theory be as good as NG for celestial mechanics: one expects from a viable theory of gravitation that it *improve* celestial mechanics as compared with NG. This requirement is usually concentrated on the demand that Mercury's residual advance in perihelion should be explained. However, the modern process of observational test of the celestial mechanics based on GR involves a general least-squares fitting of a large set of observational data by using the equations of that celestial mechanics [18]. In accordance with this process, the assessment of different celestial mechanics should not be based on God's judgment: « are Mercury's 43" predicted or not ? » but, more prosaically (and more accurately), on the comparison between the respective least-squares residuals. In doing so, one may make different comparisons based on different choices of the set of input data. *E.g.,* one may check how well-predicted is Mercury's perihelion motion, by *not* incorporating it in the input data.

The aim of this paper is to make an important theoretical step in order to make possible the numerical implementation of a celestial mechanics based on the scalar theory in an astronomical ephemeris. This step consists in obtaining the equations for the first PN corrections to the motion of the mass centers and in putting them in a tractable form. We shall base this work on foregoing results [13] that give the local equations of motion and the boundary conditions in the first PN approximation (PNA), hence *Section 2* summarizes these results. We emphasize that an « *asymptotic* » PNA is considered, in the sense that the usual method of asymptotic expansion is applied to a one-parameter family of similar gravitating systems, constructed from the data of the physical system of interest [13]; as a consequence, the gravitational field *and* the matter fields are expanded, as in Refs. 19 and 20 – in contrast with the standard PNA [4-5, 21-23], in which only the gravitational field is expanded. Moreover, we consider *extended bodies,* as is relevant to celestial mechanics [24]. To our knowledge, this is the first time that an « asymptotic » PNA is used for extended bodies. According to Asada & Futamase [25], « the theory of extended object in general relativity is still in preliminary stage for the application to realistic systems. » In *Section 3,* we describe how to properly define mass centers in order that their motion be relevant to the comparison with astronomical observations. In *Section 4,* we give the general form of the equations for the PN corrections to the motion of the mass centers. In *Section 5,* we consider the case of well-separated bodies, that introduces the maximum ratio of the size of the bodies to the distance between bodies as a small parameter $\eta$. In *Section 6,* we consider the case of « very-well-separated » bodies, in which case only terms of the lowest order in $\eta$ are conserved, together with the further relevant simplification that occurs when one assumes spherically symmetric zero-order self-fields (only when evaluating the PN corrections). In *Section 7*, we propose an algorithm for the numerical implementation of the method and we show that the Newtonian masses must be corrected. Some integrals are computed in *Appendices*.

## 2. Résumé of previous results

The theory of gravity which is considered is a scalar theory with a preferred frame E (« ether ») [8], with two space-time metrics: a flat « background metric » $\gamma^0$ and a curved « physical metric » $\gamma$, the relation between the two being defined through the scalar field [9], and with the motion being governed by an extension of Newton's second law rather than by Einstein's geodesic assumption [10]. A brief summary of the theory is given in Ref. 13 (a more detailed one is given in Ref. 14; Ref. 26 presents the complete construction of the theory, though without detailed calculations).

### *2.1 Equations for the exact fields*

The preferred frame E is assumed to be an inertial frame for the flat metric, thus there are Galilean coordinates $(x^\mu)$ for $\gamma^0$ [*i.e.,* $(\gamma^0_{\mu\nu}) = (\eta_{\mu\nu}) \equiv \text{diag}(1, -1, -1, -1)$], that are bound to the frame E. The inertial time $T \equiv x^0/c$ in the preferred frame is called the « absolute time ». In any coordinates $(y^\mu)$ bound to E and such that $y^0 = cT$, the space-time metric is given as function of the scalar field $f$ by:

$$\gamma_{00} = f, \quad \gamma_{ij} = -g_{ij}, \quad \gamma_{0i} = 0, \qquad (2.1)$$

where **g** is the spatial part of the « physical » metric $\gamma$ in the frame E, which is related to the the spatial part $\mathbf{g}^0$ of the flat metric $\gamma^0$ in the frame E as follows:

$$g_{ij} = g^0{}_{ij} + \left(\frac{1}{f} - 1\right) h_{ij}, \qquad h_{ij} \equiv \frac{f_{,i} f_{,j}}{g^{0\,kl} f_{,k} f_{,l}}. \qquad (2.2)$$

In the case that $f = 1$ in some region of space-time (which means that no gravitational field is there), the physical metric coincides with the flat metric in that region. Except for cosmological problems, the field equation [8-9] may be written as [10]:

$$\Delta f - \frac{1}{f}\left(\frac{f_{,0}}{f}\right)_{,0} = \frac{8\pi G}{c^2} \sigma \qquad (y^0 = cT), \qquad (2.3)$$

with $\Delta$ the usual Laplace operator, defined with the Euclidean metric $\mathbf{g}^0$, and where $G$ is Newton's gravitation constant, and $\sigma \equiv (T^{00})_E$ is the material mass-energy density in the ether frame. This equation is merely space-covariant. Here, we shall consider that each body is made of a perfect fluid; that is, the mass-energy tensor of matter and non-gravitational fields has the form

$$T^{\mu\nu} = (\mu^* + p/c^2)\, U^\mu U^\nu - (p/c^2)\, \gamma^{\mu\nu}, \qquad (2.4)$$

where $U^\mu = dx^\mu/ds$ is the four-velocity of the fluid, $p$ is the pressure and $(\gamma^{\mu\nu})$ is the inverse matrix of $(\gamma_{\mu\nu})$, and where $\mu^*$ is the volume density of the rest-mass plus internal energy in the proper frame, expressed in mass units,

$$\mu^* \equiv \rho^*(1 + \Pi/c^2). \qquad (2.5)$$

The dynamical equations are then given in Cartesian coordinates (such that $g^0{}_{ij} = \delta_{ij}$) by [13]:

$$\partial_T(\psi u^i) + \partial_j(\psi u^i u^j) + \Gamma^i_{jk} \psi u^j u^k + t^i{}_k \psi u^k = \psi f g^i - g^{ij} p_{,j}, \qquad (2.6)$$

$$\partial_T(\psi f - (p/c^2)) + \partial_j(\psi f u^j) = (\sigma/2) \partial_T f, \qquad (2.7)$$

where

$$u^i \equiv \frac{dx^i}{dT}, \quad g^{ij} \equiv (\mathbf{g}^{-1})_{ij}, \quad \Gamma^i_{jk} \equiv g^{il}(g_{lj,k} + g_{lk,j} - g_{jk,l}), \quad t^i{}_k \equiv \frac{1}{2} g^{ij} \partial_T g_{jk}, \quad \psi \equiv \sigma + p/(c^2 f). \quad (2.8)$$

The material mass-energy density is given by

$$\sigma \equiv (T^{00})_E = \left(\mu^* + \frac{p}{c^2}\right) \frac{\gamma_v^2}{f} - \frac{p}{c^2 f}, \qquad (2.9)$$

where $\gamma_v \equiv (1 - v^2/c^2)^{-1/2}$ is the Lorentz factor, with $v^2 \equiv g_{ij} v^i v^j$ and $v^i \equiv dx^i/dt_\mathbf{x} \equiv u^i/\sqrt{f}$. One may check that, if the scalar field $f$ can be considered equal to 1 (no gravitational field), then Eqs. (2.6) and (2.7) reduce to the dynamical equations for a perfect fluid in SR [5, 22]. However, due to Eq. (2.3), the gravitational field can hardly cancel unless there is no matter – as in GR.

## 2.2 Weak-gravitational-field limit

This is defined in the following way [13]: we consider the gravitating system of interest, S, made of perfect fluids and assumed to obey the equations of the scalar ether-theory, Eqs. (2.3), (2.6) and (2.7). We define

$$V \equiv c^2(1-f)/2, \qquad (2.10)$$

$$V_{max} \equiv \max \{V(\mathbf{x}, T=0); \mathbf{x} \in M\}, \qquad \varepsilon_0 \equiv (V_{max}/c^2)^{1/2} \equiv [(1-f_{min})/2]^{1/2} \qquad (2.11)$$

(here M is the space manifold, made of all points that are bound to the ether frame). We assume that $\varepsilon_0$ is well-defined, finite, and *small:* this is the first physical assumption about system S, it means that the gravitational field is weak. We define a one-parameter family $(S_\varepsilon)$ of gravitating systems by the fields $f^\varepsilon$, $p^\varepsilon$, $\mathbf{u}^\varepsilon$, those being defined as the solution (assumed unique for all $\mathbf{x} \in M$ and in some interval $[0, T_2(\varepsilon)]$ for $T$) of the initial-value problem

$$f^\varepsilon(\mathbf{x},0) = 1 - 2\xi^2 \frac{V(\mathbf{x},0)}{c^2}, \quad \frac{\partial f^\varepsilon}{\partial T}(\mathbf{x},0) = -\frac{2\xi^3}{c^2}\frac{\partial V}{\partial T}(\mathbf{x},0), \qquad (2.12)$$

$$p^\varepsilon(\mathbf{x}, 0) = \xi^4 p(\mathbf{x}, 0), \qquad \mathbf{u}^\varepsilon(\mathbf{x}, 0) = \xi \mathbf{u}(\mathbf{x}, 0), \qquad (2.13)$$

for equations (2.3), (2.6) and (2.7). Here $\xi = \varepsilon/\varepsilon_0$ and the state equation in system $S_\varepsilon$ is

$$\rho^{*\varepsilon} = F_\varepsilon(p^\varepsilon) \equiv \xi^2 F(\xi^{-4} p) \quad [\text{or } F_\varepsilon(p) \equiv \varepsilon^2 F_1(\varepsilon^{-4} p)], \qquad (2.14)$$

where $\rho^* = F(p)$ is the (barotropic) state equation in system S. This definition by an initial-value problem is justified by the fact that the natural boundary-value problem in the scalar ether-theory is indeed the full initial-value problem [13].

## 2.3 Asymptotic expansions of fields and equations

One expects that the family of fields just defined admits an asymptotic expansion as the small parameter $\varepsilon$ tends towards zero. More precisely, if one changes units for system $S_\varepsilon$, multiplying the starting time unit by $\varepsilon^{-1}$ and the mass unit by $\varepsilon^2$, then $\varepsilon$ becomes proportional to $1/c$ and all matter fields are of order zero with respect to $\varepsilon$. This is equivalent to say that, in invariable units, one has

$$p^\varepsilon = \text{ord}(\varepsilon^4), \qquad \mathbf{u}^\varepsilon = \text{ord}(\varepsilon), \qquad \rho^{*\varepsilon} = \text{ord}(\varepsilon^2), \qquad (2.15)$$

as in the weak-field limit of NG, and as is implied by the initial conditions (2.13) with (2.14). Since it is only $1/c^2$ that enters the field equations (2.3), (2.6) and (2.7), one states expansions in $1/c^2$ for the independent fields, adopting the varying units just described:

$$f^\varepsilon = 1 - 2 U/c^2 - 2 A/c^4 + O(\varepsilon^6), \qquad (2.16)$$
$$p^\varepsilon = p_0 + p_1/c^2 + O(\varepsilon^4), \qquad (2.17)$$
$$\mathbf{u}^\varepsilon = \mathbf{u}_0 + \mathbf{u}_1/c^2 + O(\varepsilon^4). \qquad (2.18)$$

[Theoretically, these expansions should be mathematically deduced from the definition (2.12)-(2.13).] From these expansions, one then deduces expansions for the other fields, and for the equations. Those « expanded equations » are valid in any units, provided (of course) one expresses the expanded fields in these units. As to the expansions, they take a slightly different form in invariable units [13]. However, we shall consider mainly the 1PN approximations of the fields, thus omitting the remainder terms in the expansion: *e.g.*

$$p^\varepsilon_{(1)} \equiv p_0 + p_1/c^2. \tag{2.19}$$

Those keep the same expression in any units although, in invariable units, $p_0$ is like $\varepsilon^4$ and $p_1$ like $\varepsilon^6$. Thus we shall use fixed units in this paper, unless otherwise mentioned. The order in $\varepsilon$ of an expression is easily obtained by passing through the varying units (in which the order is directly read out), but we shall leave it to the reader. Anyhow, we emphasize that, *in this work, we simply compute all expressions consistently at the first PNA, i.e. the second approximation in the expansion in powers of $\varepsilon^2$*. We also shall omit the superscript $\varepsilon$ for the 1PN approximations of the fields. Note that we are actually interested in the given system S, that corresponds to the given, small value $\varepsilon_0$ for $\varepsilon$. Finally, *we shall omit the index 0 for the zero-order coefficients in the expansions:* thus Eq. (2.19) is rewritten as

$$p_{(1)} \equiv p + p_1/c^2. \tag{2.20}$$

To avoid confusions, the exact fields, when needed, will now be denoted by the subscript « exact ». For instance, the density of rest mass in the preferred frame and with respect to the Euclidean volume measure $dV \equiv dx^1\,dx^2\,dx^3$ (in Cartesian coordinates) is defined by

$$\rho_{\text{exact}} \equiv dm_0/dV \equiv \rho^*_{\text{exact}}\,(\gamma_v)_{\text{exact}}/\sqrt{f}_{\text{exact}}\,. \tag{2.21}$$

We shall need the following relations between the expanded fields [13]:

$$\rho^* = F(p), \qquad \rho^*_1 = F'(p).p_1, \tag{2.22}$$

$$\rho = \rho^*, \qquad \rho_1 = \rho^*_1 + \rho(\mathbf{u}^2/2 + U), \tag{2.23}$$

$$\psi = \sigma = \rho = \rho^*, \qquad \sigma_1 = \rho_1 + \rho(\mathbf{u}^2/2 + \Pi + U), \qquad \psi_1 = \sigma_1 + p. \tag{2.24}$$

The expansion of the equations is as follows [13]: Eq. (2.3) is expanded to

$$\Delta U = -4\pi G\rho, \tag{2.25}$$

$$\Delta A = -4\pi G\sigma_1 + \partial^2 U/\partial T^2; \tag{2.26}$$

the energy equation (2.7) is expanded to:

$$\partial_T\rho + \partial_j(\rho u^j) = 0, \tag{2.27}$$

$$\partial_T\rho_1 + \partial_j(\rho_1 u^j + \rho u_1{}^j) = 0; \tag{2.28}$$

and the local equation of motion (2.6) is expanded to:

$$\partial_T\!\left(\rho u^i\right) + \partial_j\!\left(\rho u^i u^j\right) = \rho U_{,i} - p_{,i}, \tag{2.29}$$

$$\partial_T\left(\rho u_1^i + \psi_1 u^i\right) + \partial_j\left(\psi_1 u^i u^j + \rho u_1^i u^j + \rho u^i u_1^j\right) + {}_1\Gamma^i_{jk}\rho u^j u^k + \rho u^j \partial_T k_{ij} =$$
$$= -p_{1,i} + 2 k_{ij} p_{,j} + \psi_1 U_{,i} + \rho A_{,i} - 2U\rho U_{,i}, \tag{2.30}$$

where

$${}_1\Gamma^i_{jk} \equiv k_{ij,k} + k_{ik,j} - k_{jk,i}, \quad k_{ij} \equiv U h^1_{ij}, \quad h^1_{ij} \equiv \frac{U_{,i} U_{,j}}{U_{,k} U_{,k}}. \tag{2.31}$$

### 2.4 Boundary conditions for the expanded fields

Entering the expansions of the fields into the initial conditions (2.12)-(2.13), one gets initial conditions for the PN expansions of the matter fields [13]:

$$p(\mathbf{x}, T=0) = p_{\text{exact}}(\mathbf{x}, T=0), \quad p_1(\mathbf{x}, 0) = 0, \tag{2.32}$$

$$\rho(\mathbf{x}, 0) = F[p_{\text{exact}}(\mathbf{x}, 0)], \quad \rho_1(\mathbf{x}, 0) = [\rho(\mathbf{u}^2/2 + U)](\mathbf{x}, 0), \tag{2.33}$$

$$\mathbf{u}(\mathbf{x}, 0) = \mathbf{u}_{\text{exact}}(\mathbf{x}, 0), \quad \mathbf{u}_1(\mathbf{x}, 0) = \mathbf{0}. \tag{2.34}$$

Since the first-approximation equations: (2.25), (2.27) and (2.29), are the equations of NG, the gravitational potential $U$ plays just the role of the Newtonian potential in NG. Hence we assume spatial-boundary conditions at infinity ensuring that the solution of (2.25) is indeed the Newtonian potential associated with $\rho$:

$$(\forall T)\quad U(\mathbf{x}, T) = O(1/r) \text{ and } |\nabla U(\mathbf{x}, T)| = O(1/r^2) \text{ as } r \equiv |\mathbf{x}| \to \infty. \tag{2.35}$$

This represents a physical condition imposed on the field $f$, in addition to the requirement that $\varepsilon_0 \ll 1$ [13]. In the latter work, we proposed similar conditions for the gravitational potential $A$, but this turns out to be inappropriate. Indeed it is not $A$ but a « part » of it that must behave like a Newtonian potential [see Eqs. (4.14) and (4.15) below].

### 3. Definition of the mass centers

In the « asymptotic » scheme considered here, the decoupling between the Newtonian equations and the equations for the PN corrections is particularly clear. Hence we shall have the possibility to make stronger assumptions for the calculations of the PN corrections than for the Newtonian calculations, because the former are very small. It is in particular possible to assume spherical bodies at the level of the PN corrections to the mass centers, while keeping a more general description at the Newtonian stage. Moreover, as regards the equations for rotational motion, it is possible to content oneself with Newtonian calculations, in which case one merely has to evaluate the PN corrections to the motion of the mass centers, indeed. However, whereas in Newtonian celestial mechanics, only the distribution of the (Newtonian) mass influences the motion of the mass centers, the mass-energy equivalence changes this situation for a « relativistic » theory of gravitation: i) it is not obvious to decide which mass-energy density is relevant to the definition of the mass centers. ii) once a definition has been selected [*e.g.* one might think to the « active » energy density $\sigma$, Eqs.(2.3) and (2.9)], one has to expect that it is not only *that* density which determines the motion of the selected « mass centers ». This means that the internal structure (through its energy distribution) and the internal motion (through the corresponding kinetic energy) are now *a priori* expected to play a role. Finally, the « choice » of the energy density should be influenced by observational considerations: one should ask if the energy density selected is simply related with the emission of electromagnetic radiation detected by the telescope. It seems clear that the electromagnetic emission is very loosely correlated with the gravitational energy, because the latter is distributed in the whole space, and that instead it is strongly correlated with the rest-mass energy. In other words, it is the presence of « matter » in the usual sense, thus characterized by its rest mass, that leads to the emission of e.m. radiation. Since we are working in the preferred frame of the scalar theory, the rest-mass distribution is defined by the density (2.21), whose PN expansion involves the Newtonian density $\rho$ and the PN correction $\rho_1$.

A theoretical argument also leads to taking the rest-mass density (2.21). Indeed suppose one defines the « mass centers » as local barycenters for some density $\varphi$ in some separated domains $\omega_a$, $\omega_b$, ... [one for each body $a$, $b$, ...; the domains depend on time, of course]. Thus one defines the position **a** of the « $\varphi$-center » of body $a$ by

$$\Phi_a \mathbf{a} = \int_{\omega_a} \varphi \, \mathbf{x} \, dV, \qquad \Phi_a \equiv \int_{\omega_a} \varphi \, dV. \qquad (3.1)$$

One would wish that the « $\varphi$-mass » $\Phi_a$ be a constant, and that further one have

$$\Phi_a \dot{\mathbf{a}} = \int_{\omega_a} \varphi \, \mathbf{u}_{\text{exact}} \, dV. \qquad (3.2)$$

If a continuum moving with velocity **w** occupies just the domain $\omega_a(T)$ at any $T$, we have

$$\dot{\Phi}_a = \int_{\omega_a} \left( \frac{\partial \varphi}{\partial T} + \text{div}(\varphi \mathbf{w}) \right) dV, \qquad (3.3)$$

hence also (applying this to $\xi^i = \varphi \, x^i$):

$$\frac{d}{dT} \int_{\omega_a} \varphi \, \mathbf{x} \, dV = \int_{\omega_a} \varphi \, \mathbf{w} \, dV + \int_{\omega_a} \mathbf{x} \left( \frac{\partial \varphi}{\partial T} + \text{div}(\varphi \mathbf{w}) \right) dV. \qquad (3.4)$$

In Eqs.(3.3)-(3.4), we may alter the velocity field **w**(**x**,$T$) continuously inside $\omega_a(T)$, provided that we have $\chi_T[\omega_a(T=0)] = \omega_a(T)$, where **x**($T$) = $\chi_T$(**X**) are integral lines of the vector field **w** (such that d**x**/d$T$ = **w** on these lines; **X** is the position at time $T=0$). Moreover, in the special case that $\varphi(\mathbf{x},T) = 0$ on the boundary $\partial \omega_a(T)$, an application of the divergence theorem shows that (3.3)-(3.4) hold true for an arbitrary vector field **w**(**x**,$T$). Hence, Eqs.(3.1)-(3.2) are certainly satisfied if $\varphi$ obeys the usual continuity equation with the exact velocity $\mathbf{u}_{\text{exact}}$, and there is hope that one may substitute some approximate velocity **w** for $\mathbf{u}_{\text{exact}}$, provided $\varphi$ and **w** satisfy the usual continuity equation. In the scalar theory, there is an exact conservation equation for the total energy, including the gravitational energy [10]. But this conservation equation [obtained by rewriting the r.h.s. of (2.7) using the field equation (2.3)] is not the usual continuity equation since it also involves gravitational energy density and flux. Only the rest-mass density obeys the usual continuity equation, though it obeys this equation approximately [see Eq.(3.11) below].

Thus, we define the exact mass center through the rest-mass density $\rho_{\text{exact}}$:

$$M_a^{\text{exact}} \mathbf{a}_{\text{exact}} = \int_{\omega_a} \rho_{\text{exact}} \mathbf{x}\, dV, \qquad M_a^{\text{exact}} \equiv \int_{\omega_a} \rho_{\text{exact}}\, dV. \qquad (3.5)$$

At the (first) PNA, $\rho_{\text{exact}}$ is approximated as $\rho_{\text{exact}} = \rho_{(1)} [1+ \text{O}(\varepsilon^4)]$, with [*cf.* (2.22), (2.23)]

$$\rho_{(1)} = \rho + \rho_1/c^2, \qquad \rho_1 = (\mathbf{u}^2/2 + U)F(p) + p_1 F'(p) \qquad (3.6)$$

Thus, the mass and the mass center are approximated by

$$M_a^{(1)} = M_a + M_a^1 / c^2, \quad M_a \equiv \int_{\omega_a} \rho\, dV, \quad M_a^1 \equiv \int_{\omega_a} \rho_1\, dV, \qquad (3.7)$$

$$M_a^{(1)} \mathbf{a}_{(1)} = \int_{\omega_a} \rho_{(1)} \mathbf{x}\, dV = M_a \mathbf{a} + M_a^1 \mathbf{a}_1 / c^2, \qquad (3.8)$$

with

$$M_a \mathbf{a} = \int_{\omega_a} \rho \mathbf{x}\, dV, \qquad M_a^1 \mathbf{a}_1 = \int_{\omega_a} \rho_1 \mathbf{x}\, dV. \qquad (3.9)$$

Note that $M_a$ and $\mathbf{a}$ are the Newtonian mass and center of mass. Now let us show that $M_a$ and $M_a^1$ are (nearly) constant. We first have to precise the definition of the domain $\omega_a$ occupied by body $a$ (this does not appear to be done in the literature on relativistic celestial mechanics). In the real world, there is no sharp separation between different celestial bodies, for a perfect *vacuum* does not exist. This is compatible with different barotropic state equations for different bodies, provided the state equations coincide at very low pressure. We define the domain $\omega_a(T = 0)$ occupied by body $a$ at the initial time by a threshold pressure, thus $q(\mathbf{X}) \equiv p(\mathbf{X},T = 0) > p_{\min}^a$ if $\mathbf{X}$ belongs to $\omega_a(T = 0)$. Then we follow $\omega_a$ with the PN velocity as the velocity $\mathbf{w}$ :

$$\mathbf{w} = \mathbf{u}_{(1)} \equiv \mathbf{u} + \mathbf{u}_1/c^2. \qquad (3.10)$$

Now we have by (2.27), (2.28) and (2.15):

$$\frac{\partial \rho_{(1)}}{\partial T} + \text{div}\left(\rho_{(1)} \mathbf{u}_{(1)}\right) = \frac{1}{c^4} \text{div}\left(\rho_1 \mathbf{u}_1\right) = \text{O}(\varepsilon^7), \qquad (3.11)$$

from where it follows by (3.3) and (3.7)$_1$ that $M_a^{(1)}$ is conserved at the first PNA:

$$dM_a^{(1)}/dT = \text{O}(\varepsilon^7), \qquad (dM_a^{(1)}/dT)/M_a^{(1)} = \text{O}(\varepsilon^5). \qquad (3.12)$$

(Actually, a global mass like this is likely to be conserved with an *extremely* good approximation in the scalar theory [27].) Moreover, from the continuity equation (2.27), we get using Eqs.(3.3), (3.7) and (3.10), then (2.15):

$$dM_a/dT = \int_{\partial \omega_a} \rho\, \mathbf{u}_1 \cdot \mathbf{n}/c^2\, dS = \text{O}(\varepsilon^5). \qquad (3.13)$$

Note that $\rho$ follows the Newtonian motion that has velocity $\mathbf{u} \neq \mathbf{w}$. For instance, in the case of Mercury, the « Newtonian » body will be, after one century, at 43'' angular distance on Mercury's orbit from the « PN » body (if the PNA of the theory is good enough to reproduce the observed residual advance). This corresponds to some $12.10^3$ km, to be compared with the 2437 km of Mercury's « official » equatorial radius. However, by taking $p_{\min}^{\text{Mercury}}$ small enough, the radius of Mercury shall be several $10^4$ km, say, and yet this « enlarged Mercury » will remain at a comfortable distance from other celestial bodies. With this definition, the « Newtonian » position of the body and the « PN » one, though at the same distance from one another (roughly $10^4$ km after one century, in the case of Mercury), will both contain virtually all the matter of the real body, up to a negligible amount. In other words, by taking $p_{\min}^a$ small enough, we may assume that $\rho$ and $\rho_1$ are negligible on $\partial\omega_a$ so that, in particular, the rate given by (3.13) [which rate is $O(\varepsilon^5)$ independently of this « enlargement »] is utterly negligible. By (3.12) and (3.7)$_1$, $dM_a^1/dT$ is $O(\varepsilon^5)$ also but, actually, from (2.27), we get using Eqs.(3.3), (3.7)$_3$ and (3.10):

$$dM_a^1/dT = \int_{\partial\omega_a} (\rho_1 \mathbf{u}_1/c^2 - \rho \mathbf{u}_1).\mathbf{n}\, dS, \qquad (3.14)$$

which is utterly negligible with our definition of « enlarged » bodies. But, after some time (may be of the order of a century), one should «reinitialize» the problem. We get from (3.7)-(3.9):

$$\delta\mathbf{a} \equiv \mathbf{a}_{(1)} - \mathbf{a} = \frac{1}{c^2}\left[\frac{M_a^1}{M_a}(\mathbf{a}_1 - \mathbf{a})\right] + O(\varepsilon^4). \qquad (3.15)$$

As to $M_a^1$, we obtain by using (2.33)$_2$ in the definition (3.7)$_3$:

$$M_a^1 = M_a^1(T=0) = \left[\int_{\omega_a} \rho(\mathbf{u}^2/2 + U)\, dV\right]_{T=0}. \qquad (3.16)$$

We may express the velocity of the PN mass center (3.8) as an average of the velocity inside the body $a$. Applying (3.4) with $\mathbf{w} = \mathbf{u}_{(1)}$, and first with $\varphi = \rho_{(1)}$, then with $\varphi = \rho$, we get by using first (3.11), then (2.27):

$$M_a^{(1)}\dot{\mathbf{a}}_{(1)} = \int_{\omega_a} \rho_{(1)} \mathbf{u}_{(1)}\, dV + O(\varepsilon^7), \qquad (3.17)$$

$$M_a\, \dot{\mathbf{a}} = \int_{\omega_a} \{\rho\, \mathbf{u}_{(1)} + \mathbf{x}\, \text{div}[\rho(\mathbf{u}_{(1)} - \mathbf{u})]\}\, dV = \int_{\omega_a} \rho\, \mathbf{u}\, dV + \int_{\partial\omega_a} \mathbf{x}\rho(\mathbf{u}_{(1)} - \mathbf{u}).\mathbf{n}\, dS. \qquad (3.18)$$

The surface integral is merely $O(\varepsilon^5)$ by (2.15), but it is utterly negligible with our definition of « enlarged » bodies enclosing both the perturbed and the Newtonian position of the massive part. Thus we may consider that

$$M_a\, \dot{\mathbf{a}} = \int_{\omega_a} \rho\, \mathbf{u}\, dV\ [= \text{ord}(\varepsilon^3)]. \qquad (3.19)$$

Therefore, we have

$$M_a^{(1)}\dot{\mathbf{a}}_{(1)} = M_a\, \dot{\mathbf{a}} + M_a^1\, \dot{\mathbf{a}}_1/c^2, \qquad (3.20)$$

where the Newtonian term is given by (3.19), and with [using also (3.17)]

$$M_a^1 \, \dot{\mathbf{a}}_1/c^2 = \int_{\omega_a} [(\rho_1 \mathbf{u} + \rho \mathbf{u}_1)/c^2] dV + O(\varepsilon^7). \qquad (3.21)$$

## 4. Equations of motion for the mass centers: i) General form of the PN equations

To find the equations that govern the motion of the mass centers, we simply integrate the local equations of motion, using the following result [deduced from Eq.(3.3)]:

$$\partial_T z^i + \partial_j (z^i v^j) = f^i \;\Rightarrow\; \frac{d}{dT}\int_{\omega_a} \mathbf{z}\, dV = \int_{\omega_a} \mathbf{f}\, dV + \int_{\partial \omega_a} \mathbf{z}[(\mathbf{w}-\mathbf{v}).\mathbf{n}]\, dS. \qquad (4.1)$$

If $\omega_a$ moves with the exact velocity: $\mathbf{w} = \mathbf{v} = \mathbf{u}_{\text{exact}}$, then, with $\mathbf{z} = \psi\mathbf{u}$, we obtain from (2.6):

$$\frac{d}{dT}\left(\int_{\omega_a} \psi u^i \, dV\right) = \int_{\omega_a}\left(\psi f g^i - g^{ij} p_{,j} - \Gamma^i_{jk}\psi u^j u^k - t^i{}_k \psi u^k\right) dV, \qquad (4.2)$$

in which all fields are the exact fields. Coming back to the PNA and to the notations used since Eq. (2.20), $\omega_a$ moves with the PN velocity: $\mathbf{w} = \mathbf{v} = \mathbf{u}_{(1)}$, and we take $\mathbf{z} = \psi_{(1)} \mathbf{u}_{(1)}$ [where $\psi_{(1)} \equiv \rho + \psi_1/c^2$ is the 1PN approximation of $\psi$]. Using the PN equations of motion (2.29) and (2.30), we get just the PN expansion of (4.2):

$$\frac{d}{dT}\left\{\int_{\omega_a} \left[\rho u^i + \left(\rho u_1^i + \psi_1 u^i\right)/c^2\right] dV\right\} = \int_{\omega_a}\left(f^i + f_1^i/c^2\right) dV, \qquad (4.3)$$

with

$$f^i \equiv \rho\, U_{,i}, \qquad (4.4)$$

$$f_1^i \equiv 2\, k_{ij}\, p_{,j} + \psi_1\, U_{,i} + \rho\, A_{,i} - 2U\rho U_{,i} - {}_1\Gamma^i_{jk}\rho u^j u^k - \rho u^j \partial_T k_{ij}. \qquad (4.5)$$

(The terms $p_{,i}$ and $p_{1,i}$ have been removed from $f^i$ and $f_1^i$ respectively, by using the divergence theorem and the definition of « enlarged » bodies, see Sect. 3.) As is obvious in the varying units (see §2.3), we may identify the powers of $1/c^2$ in Eq. (4.3). Owing to (2.24), we may then insert Eqs. (3.19) and (3.21) respectively. We obtain thus

$$M_a \ddot{a}^i = \int_{\omega_a} \rho\, U_{,i}\, dV, \qquad (4.6)$$

$$M_a^1 \ddot{a}_1^i + \dot{I}^{ai} = \int_{\omega_a} f_1^i\, dV, \qquad (4.7)$$

where

$$I^{ai} \equiv \int_{\omega_a} [p + \rho(\mathbf{u}^2/2 + \Pi + U)] u^i\, dV. \qquad (4.8)$$

To simplify the calculation of the PN correction (4.7) to the motion, we rewrite (4.7) as

$$M_a^1 \ddot{a}_1^i + \dot{I}^{ai} = J^{ai} + K^{ai}, \qquad (4.9)$$

with

$$K^{ai} \equiv \int_{\omega_a} [2\, k_{ij}\, p_{,j} + p\, U_{,i} - 2U\rho U_{,i} - {}_1\Gamma^i_{jk}\, \rho u^j u^k - \rho u^j \partial_T k_{ij}]\, dV \qquad (4.10)$$

and

$$J^{ai} \equiv \int_{\omega_a} (\sigma_1 U_{,i} + \rho A_{,i})\, dV. \qquad (4.11)$$

Equation (4.9) is the *equation for the PN correction to the motion of the mass center of body (a)*. Note that $I^{ai}$ and $K^{ai}$ depend only on the first-approximation (Newtonian) fields. To evaluate $J^{ai}$, we adapt Fock's method [5, §§ 75-77]. If we define $U^* \equiv U + A/c^2$ and $\tau \equiv \sigma_{(1)} \equiv \rho + \sigma_1/c^2$, we have $A\sigma_1/c^4 = O(\varepsilon^8)$ by (2.15), hence

$$\int_{\omega_a} \tau U^*_{,i}\, dV = \int_{\omega_a} \rho\, U_{,i}\, dV + J^{ai}/c^2 + O(\varepsilon^8) = \int_{\omega_a} \rho\, U^{(a)}_{,i}\, dV + J^{ai}/c^2 + O(\varepsilon^8)\ [=\mathrm{ord}(\varepsilon^4)]. \quad (4.12)$$

Here and in the following, we use Fock's decomposition of any field $Z$, integral of some density $\theta$ vanishing outside the bodies, into « self » and « external » parts $z_a$ and $Z^{(a)}$:

$$Z \equiv \int \theta\, dV = z_a + Z^{(a)}, \quad z_a \equiv \int_{\omega_a} \theta\, dV, \quad Z^{(a)} \equiv \sum_{b \neq a} \int_{\omega_b} \theta\, dV. \qquad (4.13)$$

The second equality in (4.12) is due to the fact that, owing to Poisson's equation (2.25) and the boundary conditions (2.35), $U$ is the Newtonian potential associated with $\rho$ (let us denote this by $U$ = N.P.$[\rho]$), hence the integral of the self-force $\rho\, u_{a,i}$ vanishes. Defining

$$W(\mathbf{X}, T) \equiv \int G R\, \rho(\mathbf{x}, T)\, dV(\mathbf{x})/2, \qquad R \equiv |\mathbf{X} - \mathbf{x}|, \qquad (4.14)$$

we have $\Delta W = U$. Thus (2.26) may be rewritten as

$$\Delta B = -4\pi G \sigma_1, \qquad B \equiv A - \partial^2 W/\partial T^2.$$

The source of $B$, thus $\sigma_1$, vanishes outside the bodies − contrary to the source of $A$. Hence it is natural to expect that $B$ should be the Newtonian potential associated with $\sigma_1$. Therefore we impose on $B$ the same conditions as Eqs. (2.35)$_{1\text{-}2}$ for $U$, which ensures that indeed $B$ = N.P.$[\sigma_1]$. {But $\partial^2 W/\partial T^2$ does *not* obey Eqs. (2.35) [see Eq. (B6) below], thus $A$ does *not* behave like a Newtonian potential.} Thus we get:

$$A(\mathbf{X},T) = B + \frac{\partial^2 W}{\partial T^2} = \int \frac{G}{R} \sigma_1(\mathbf{x},T)\, dV(\mathbf{x}) + \frac{\partial^2 W}{\partial T^2}. \qquad (4.15)$$

Hence $U^* = U^\dagger + (\partial^2 W/\partial T^2)/c^2$ with $U^\dagger \equiv$ N.P.$[\tau]$. Therefore, we have as for (4.12)$_2$:

$$\int_{\omega_a} \tau U^*_{,i}\, dV = \int_{\omega_a} \tau\, U^{\dagger(a)}_{,i}\, dV + \int_{\omega_a} \tau\, (\partial^3 W/\partial x^i \partial T^2)/c^2\, dV$$

$$= \int_{\omega_a} \tau\, U^{\dagger(a)}_{,i}\, dV + \int_{\omega_a} \rho\, (\partial^3 W/\partial x^i \partial T^2)/c^2\, dV + O(\varepsilon^8). \qquad (4.16)$$

By the definitions of $\tau$ and $U^\dagger$, we have

$$U^\dagger = U + B/c^2, \qquad B = \mathrm{N.P.}[\sigma_1]. \qquad (4.17)$$

From (4.12) and (4.16), it thus follows that

$$J^{ai} = \int_{\omega_a} \sigma_1 U^{(a)}{}_{,i}\, dV + \int_{\omega_a} \rho B^{(a)}{}_{,i}\, dV + \int_{\omega_a} \rho\, (\partial^3 W/\partial x^i\, \partial T^2)\, dV + O(\varepsilon^8). \tag{4.18}$$

## 5. PN corrections for the mass centers: ii) well-separated rigidly-moving bodies

Since the two first integrals in Eq. (4.18) involve only external gravitational fields, we may use « multipole expansions » : for $\mathbf{x} \in \omega_b$ and $\mathbf{X} \in \omega_a$ with $a \neq b$,

$$\frac{1}{R} \equiv \frac{1}{|\mathbf{X} - \mathbf{x}|} = \frac{1}{|\mathbf{X} - \mathbf{b}|} + \frac{(\mathbf{x} - \mathbf{b}) \cdot (\mathbf{X} - \mathbf{b})}{|\mathbf{X} - \mathbf{b}|^3} + O\left(\frac{\eta^2}{R}\right), \qquad \eta \equiv \mathrm{Sup}_{a \neq b}(r_b / |\mathbf{a} - \mathbf{b}|) \tag{5.1}$$

[$r_b$ is the radius of body ($b$)]. However, the source of $B$ includes a self-field: we have

$$\sigma_1 = \rho_1 + \rho\, (\mathbf{u}^2/2 + \Pi + U) = \varphi_a + \rho U^{(a)}, \qquad \varphi_a \equiv \rho_1 + \rho\, (\mathbf{u}^2/2 + \Pi + u_a) \tag{5.2}$$

whence

$$B^{(a)}/G \equiv \Sigma_{b \neq a} \int_{\omega_b} \sigma_1/R\, dV = \Sigma_{b \neq a} \left( \int_{\omega_b} \varphi_b/R\, dV + \int_{\omega_b} \rho U^{(b)}/R\, dV \right) \tag{5.3}$$

Using this, it is shown in Appendix A that, for the case where *each body undergoes a rigid motion at the Newtonian approximation:*

$$u^i = \dot{a}^i + \Omega^{(a)}{}_{ji}\, (x^j - a^j) \text{ inside } \omega_a, \quad (\Omega^{(a)}{}_{ji} + \Omega^{(a)}{}_{ij} = 0) \tag{5.4}$$

and, as does Fock, *neglecting* $O(\eta^2)$, *but not* $O(\eta)$, *with respect to* $O(\eta^0)$, one has

$$\frac{B^{(a)}(\mathbf{X})}{G} = \sum_{b \neq a} \left\{ \frac{M_b[\tfrac{1}{2}\dot{\mathbf{b}}^2 + U^{(b)}(\mathbf{b})] + \varsigma_b + M_b^1}{|\mathbf{X} - \mathbf{b}|} + \frac{[I^{(b)}_{ij}\Omega^{(b)}_{jk}\dot{b}^k + \varsigma_{bi} + M_b^1(b_1^i - b^i)](X^i - b^i)}{|\mathbf{X} - \mathbf{b}|^3} \right\},$$

$$(\mathbf{X} \in \omega_a); \tag{5.5}$$

$$\int_{\omega_a} \rho B^{(a)}{}_{,i}\, dV = M_a \left( \frac{\partial B^{(a)}}{\partial X^i} \right)_{\mathbf{X} = \mathbf{a}}; \tag{5.6}$$

$$\int_{\omega_a} \sigma_1 U^{(a)}{}_{,i}\, dV = \left( M_a[\tfrac{1}{2}\dot{\mathbf{a}}^2 + U^{(a)}(\mathbf{a})] + \varsigma_a + M_a^1 \right) \frac{\partial U^{(a)}}{\partial X^i}\bigg|_{\mathbf{X} = \mathbf{a}} +$$

$$+ [I^{(a)}_{jl}\Omega^{(a)}_{lk}\dot{a}^k + \varsigma_{aj} + M_a^1(a_1^j - a^j)] \frac{\partial^2 U^{(a)}}{\partial X^i \partial X^j}\bigg|_{\mathbf{X} = \mathbf{a}}. \tag{5.7}$$

In Eqs.(5.5) and (5.7), $I^{(a)}{}_{jk}$ is the inertia tensor:

$$I^{(a)}{}_{jk} \equiv \int_{\omega_a} \rho (x^j - a^j)(x^k - a^k)\, dV, \tag{5.8}$$

and

$$\zeta_a \equiv (8T_a + 11\varepsilon_a)/3, \qquad \zeta_{ai} \equiv (5T_{ai} + 8\varepsilon_{ai} - \eta_{ai})/2, \tag{5.9}$$

where $T_a$, $\varepsilon_a$, $T_{ai}$, $\varepsilon_{ai}$ and $\eta_{ai}$ are introduced by Fock [5, §74] and depend only on the rotational and self-potential energies and their distribution:

$$T_a \equiv \int_{\omega_a} \rho \Omega_a\, dV = \Omega^{(a)}{}_{ik}\, \Omega^{(a)}{}_{jk}\, I^{(a)}{}_{ij}/2, \qquad \Omega_a(\mathbf{x}) \equiv \Omega^{(a)}{}_{ik}\, \Omega^{(a)}{}_{jk}\, (x^i - a^i)(x^j - a^j)/2, \tag{5.10}$$

$$T_{ai} \equiv \int_{\omega_a} \rho \Omega_a\, (x^i - a^i)\, dV, \tag{5.11}$$

$$\varepsilon_a \equiv \int_{\omega_a} \rho\, u_a\, dV/2 = \int_{\text{space}} (\text{grad } u_a)^2\, dV/8\pi G = B^{(a)}{}_{kk}, \tag{5.12}$$

$$\varepsilon_{ai} \equiv \int_{\omega_a} \rho\, u_a\, (x^i - a^i)\, dV/2 = \int_{\text{space}} (\text{grad } u_a)^2\, (x^i - a^i)\, dV/8\pi G = B^{(a)}{}_{i\,kk}, \tag{5.13}$$

$$\eta_{ai} \equiv (B^{(a)}{}_{k\,ik} + B^{(a)}{}_{i\,kk})/2, \tag{5.14}$$

with

$$B^{(a)}{}_{kl} \equiv \int_{\text{space}} [\delta_{kl}(\text{grad } u_a)^2/2 - u_{a,k}\, u_{a,l}]\, dV/4\pi G \tag{5.15}$$

and

$$B^{(a)}{}_{i\,kl} \equiv \int_{\text{space}} [\delta_{kl}(\text{grad } u_a)^2/2 - u_{a,k}\, u_{a,l}]\, (x^i - a^i)\, dV/4\pi G. \tag{5.16}$$

Hence, we get finally from (4.18) and (5.5)-(5.7):

$$J^{ai} = GM_a\, \frac{\partial}{\partial X^i}\bigg|_{\mathbf{X}=\mathbf{a}} \sum_{b \neq a} \left[ \frac{\alpha_b}{|\mathbf{X}-\mathbf{b}|} + \left(\beta_{bj} + M_b^1 b_1^j\right)\frac{X^j - b^j}{|\mathbf{X}-\mathbf{b}|^3} \right] + \alpha_a \frac{\partial U^{(a)}}{\partial a^i} + \left(\beta_{ai} + M_a^1 a_1^i\right)\frac{\partial^2 U^{(a)}}{\partial a^i \partial a^j} + L^{ai} \tag{5.17}$$

where the notation $\partial/\partial a^i$ is used for $(\partial/\partial X^i)_{\mathbf{X}=\mathbf{a}}$ when no confusion may occur, and with

$$\alpha_a \equiv M_a[\dot{\mathbf{a}}^2/2 + U^{(a)}(\mathbf{a})] + \zeta_a + M_a^1, \qquad \beta_{ai} \equiv I^{(a)}{}_{ij}\, \Omega^{(a)}{}_{jk}\, \dot{a}^k + \zeta_{ai} - M_a^1 a^i, \tag{5.18}$$

$$L^{ai} \equiv \int_{\omega_a} \rho\, (\partial^3 W/\partial x^i \partial T^2)\, dV. \tag{5.19}$$

Thus, if the bodies are « *well-separated* » [$\eta \ll 1$ in Eq. (5.1)] and *rigidly moving* in the sense of Eq. (5.4), then the equation (4.9) for the PN correction to the mass-centers motion depends only on Newtonian quantities, *and, linearly,* on the PN corrections $\mathbf{a}_1$ to the positions of the mass centers and the (constant) PN corrections $M_a^1$ to the masses. Note that the $\mathbf{a}_1$'s are precisely the unknowns of Eq. (4.9). Clearly, the two assumptions above are well-justified in celestial mechanics, especially in the solar system (and the more so as we are here calculating PN *corrections*). The Newtonian quantities involved are: $M_a$, $\mathbf{a}$, $\dot{\mathbf{a}}$, $I^{(a)}{}_{jl}$, $\Omega^{(a)}{}_{ji}$, $\zeta_a$ and $\zeta_{ai}$, plus $I^{ai}$ [Eq.(4.8)], $K^{ai}$ [Eq.(4.10)], and $L^{ai}$. Theoretically, the Newtonian fields, hence all these quantities, should be known at the stage of PN correction, in which case the resolution of (4.9) would be very easy. However, it remains to obtain explicit expressions for $I^{ai}$, $K^{ai}$ and $L^{ai}$, by using the separation into

« self » and « external » fields; a first step for this is done in Appendix A, leading to Eqs. (A10)-(A11), (A13), (A21) and (A24). It appears from these equations that explicit expressions seem unlikely to be obtainable unless further simplifications are introduced.

### 6. PN corrections for the mass centers: iii) very-well-separated spherical bodies

*6.1 Introduction of the « very good separation » and « spherical symmetry » assumptions*

In the solar system, the separation between bodies is much larger than their dimensions: *e.g.* the ratio (Earth radius)/(Earth-Sun distance) is $\eta_{E\text{-}S} = 4.10^{-5}$, even the ratio (Sun radius)/(Earth-Sun distance) is still $\eta_{S\text{-}E} = 5.10^{-3}$. Now, in Eqs. (A11) and (A24) for the « external » part of the PN corrections, the second term on the r.h.s. is systematically of order $\eta$ times the first term. More precisely, in (A11), we have

$$M_a \dot{a}^i U^{(a)}(\mathbf{a}) \approx \sum_{b \neq a} G M_a M_b \dot{a} / |\mathbf{a} - \mathbf{b}|, \tag{6.1}$$

$$I^{(a)}{}_{jk} \Omega^{(a)}{}_{ki} U^{(a)}{}_{,j}(\mathbf{a}) \approx \sum_{b \neq a} M_a r_a^2 \times (V_{\text{rot}}^{(a)}/r_a) \times GM_b / |\mathbf{a} - \mathbf{b}|^2, \tag{6.2}$$

(with $V_{\text{rot}}^{(a)} = \Omega^{(a)} r_a$ the linear rotation velocity), hence the ratio is of order

$$\rho_{ab} \approx \eta_{ab} V_{\text{rot}}^{(a)}/\dot{a}. \tag{6.3}$$

Thus, $\eta_{ab}$ is multiplied by the other parameter

$$\chi^{(a)} \equiv V_{\text{rot}}^{(a)}/\dot{a}, \tag{6.4}$$

for which estimates shall be given below. In the same way, on the r.h.s. of (A24), we have two sums. The two terms in the first sum are of the same order of magnitude. The first term in the second sum is of order $(\chi^{(a)})^2$ times the former terms, and the second one is of order $\rho_{ba}$ times those same terms. Now recall that the equations for PN corrections, as well as the exact equations of the scalar theory [and in contrast with the first-approximation (Newtonian) equations], must be written in the preferred reference frame. From what we know about intra- and inter-galactical relative velocities, as also if we make the working assumption that the cosmic microwave background is at rest in that « ether », we expect that the velocity of the solar system through the « ether » should be something like 300 km/s, which would imply that the absolute velocities of the bodies would all have this same order of magnitude. Hence the parameters $\chi^{(a)}$, instead of being of the order of unity (for Jupiter or Saturn) to $10^{-2}$ (*e.g.* for the Earth) in the solar system, as if one would take the orbital velocities in Copernic's reference frame, should be of the order 1/30 to $10^{-3}$. Thus, the ratios $\rho_{ab}$ should be of the order $10^{-4}$ to $10^{-8}$, *i.e.* very small, and $(\chi^{(a)})^2$ should be of the order $10^{-3}$ to $10^{-6}$, *i.e.* quite small also. Yet it seems dangerous to assume in advance the absolute velocity, hence *we shall not neglect the self-rotation of the bodies,* since for giant planets it contributes a kinetic energy of the same order of magnitude as does the orbital motion.

However, we shall use the simplifying assumption of *spherical symmetry* for all bodies in the system, in order to get tractable calculations for $K^{ai}$. Obviously, this is not too far from the reality as regards the Sun, planets and satellites, but it is difficult to assess directly the numerical error which is done with this assumption. What we know is the geometrical flattening at the poles, $f_a \equiv (r_a - r'_a)/r_a$ (with $r_a$ and $r'_a$ the equatorial and polar radius; the flattening is due to the self-rotation). This is negligible for Venus and Mercury, very small for the Earth and March (0.003 and 0.005 respectively), but it is as large as 0.06 for Jupiter and 0.1 for Saturn. As to the Sun, there has been a controversy: according to Dicke & Goldenberg [28], one would have $f_S \approx 43 \times 10^{-3}/959 \approx 4.5 \times 10^{-5}$, whereas Hill *et al.* [29] found $f_S \approx 9 \times 10^{-3}/959 \approx 9.4 \times 10^{-6}$ [4]. The controversy arose because the value found by Dicke & Goldenberg would have contributed some 4" by century to Mercury's perihelion shift, *i.e.* a 10% fraction of the 43" « residual » shift. Of interest to us is the fact that 4" make a fraction $4/(538 \times 10^6) \approx 7.4 \times 10^{-9}$ of the Newtonian main term (here the orbital angular displacement after one century). Note that this 4" contribution is indeed the *Newtonian* (first-approximation) correction to the orbital motion of Mercury, due to the oblateness of the Sun (according to its measurement by Dicke & Goldenberg). Let us provisionally admit that the effects of oblateness depend linearly on the oblateness $f_a$ (which is true for small oblatenesses) and that moreover they imply the same *relative* error, whether in any Newtonian term or in the corresponding PN correction (which is a rough conjecture). Then we expect from this example that even an oblateness of 0.06 (that of Jupiter) would give a relative contribution of merely some $0.06 \times 7.4 \times 10^{-9}/(4.5 \times 10^{-5}) \approx 10^{-5}$ to a given PN correction. Of course, this is a very rough estimate, yet it indicates that the assumption of spherical bodies should involve only a small error when calculating the PN corrections in the solar system. Anyway, this is the assumption made in GR also when calculating the PN corrections in the solar system [4, 18, 22, 23]. Thus, *we shall assume spherically symmetric bodies henceforth.* This assumption will concern merely *the self-fields of the first approximation, as used for the calculations of second-approximation corrections.* Therefore, the incompatibility between exact spherical symmetry and the presence of self-rotation is not relevant here. Moreover, *we shall retain only the first non-zero term in the expansion with respect to the separation parameter $\eta$.* This is consistent with the orders of magnitude involved: if one would wish to account for the next order in $\eta$, one should also remove the sphericity assumption. It turns out that, when one encounters two successive powers of $\eta$, the ratio of the second term to the first one is often of order $\eta \chi^{(a)}$. Recall that $\chi^{(a)}$ is small in Copernic's frame, *except for the giant planets,* and that $\chi^{(a)}$ *should be* small for all bodies in the preferred frame. We conclude that the two foregoing assumptions involve a relative error which is almost certainly smaller than $10^{-2}$, and which is actually likely to be smaller than $10^{-3}$, when calculating the PN corrections in the solar system. We emphasize that these two assumptions, together with the *additional neglect of the self-rotation* (which we shall *not* impose), are made when calculating PN corrections for the solar system *in GR,* since the Einstein-Infeld-Hoffmann equations (based on point masses) are very generally used for practical calculations [18]. Since, in GR, Copernic's frame is relevant, the neglect of the self-rotation of giant planets would need some justification.

*6.2 Simplifications resulting from spherical symmetry and « very good separation »*

In (A11), the second term on the r.h.s., being $O(\eta)$ times the first term, now disappears. In (A24), the third and fourth terms disappear also, because the inertia tensor of a spherical body is

$$I^{(a)}{}_{ij} = \gamma_a \, \delta_{ij}, \qquad \gamma_a \equiv (4\pi/3) \int_0^{r_a} r^4 \rho(r) \, dr = \int_{(a)} r^2 \rho \, dV/3, \qquad (6.5)$$

which is a constant for the case of rigid motion. Passing to Eqs. (A10) and (A21) that are concerned with the « self » part of the PN corrections, we note that the ratio

$$\dot{a}^k \Omega^{(a)}{}_{lk} I^{(a)}{}_{jl} \Omega^{(a)}{}_{ji}/(M_a \dot{\mathbf{a}}^2 \dot{a}^i),$$

which is of the order $(\chi^{(a)})^2$, is like the ratio of the rotational to translational kinetic energies, hence the numerator term above should be conserved in (A10); with (6.5), it becomes:

$$\dot{a}^k \Omega^{(a)}{}_{lk} I^{(a)}{}_{jl} \Omega^{(a)}{}_{ji} = \gamma_a \Omega^{(a)}{}_{ji} \Omega^{(a)}{}_{jk} \dot{a}^k. \tag{6.6}$$

In (A21), the ratio

$$\varepsilon_{aj} \Omega^{(a)}{}_{ji}/(\varepsilon_a \dot{a}^i)$$

which is *a priori* (independently of any shape factor) of order $\chi^{(a)}$, cancels in the case of spherical symmetry, currently considered. In fact, integrals like $\varepsilon_{aj}$ and $T_{aj}$ [Eqs. (5.13) and (5.11)] are exactly zero in the more general case of « axial plus planar » symmetry where the field to be barycentered ($\rho u_a$ or $\rho \Omega_a$) is invariant by rotation around some axis and by symmetry with respect to some plane which is perpendicular to that axis – the axis intersecting the plane at **a**. Note that this is *very nearly* the case for most massive celestial bodies. Moreover, for spherical symmetry, we have [using Eq. (C4) in the definition (5.15)]:

$$B^{(a)}{}_{ik} = \delta_{ik} \int_0^\infty 4\pi [(1/2) - (1/3)] r^2 (du_a/dr)^2 \, dr/4\pi G = \delta_{ik} \varepsilon_a/3 \tag{6.7}$$

and

$$B^{(a)}{}_{i\,kl} = \int_0^\infty [\delta_{kl}/2 - n_k n_l] r^3 (du_a/dr)^2 \, n_i \, dr \, d\omega /4\pi G = 0, \tag{6.8}$$

because

$$\int_{\text{sphere}} n_i \, d\omega = 0, \qquad \int_{\text{sphere}} n_i n_j n_k \, d\omega = 0. \tag{6.9}$$

Compiling Eqs. (A10)-(A11), (A21) and (A24), we obtain under our assumptions:

$$I^{ai} = [M_a \dot{\mathbf{a}}^2/2 + 2T_a + 4\varepsilon_a] \dot{a}^i + \gamma_a \Omega^{(a)}{}_{ji} \Omega^{(a)}{}_{jk} \dot{a}^k + M_a \dot{a}^i U^{(a)}(\mathbf{a}), \tag{6.10}$$

$$L^{ai} = -(2/3)\varepsilon_a \ddot{a}^i - \frac{GM_a}{2} \sum_{b \neq a} M_b \left( \ddot{b}^k \frac{\partial^2 |\mathbf{a}-\mathbf{b}|}{\partial a^i \partial a^k} - \dot{b}^k \dot{b}^j \frac{\partial^3 |\mathbf{a}-\mathbf{b}|}{\partial a^i \partial a^k \partial a^j} \right). \tag{6.11}$$

To the same approximation, *i.e.* neglecting $\eta$ as compared with 1, we may also rewrite (5.17) as

$$J^{ai} - L^{ai} = \alpha_a \frac{\partial U^{(a)}}{\partial a^i} + GM_a \frac{\partial}{\partial X^i} \bigg|_{\mathbf{X}=\mathbf{a}} \sum_{b \neq a} \frac{\alpha_b}{|\mathbf{X}-\mathbf{b}|} . \tag{6.12}$$

(Note that $\zeta_{ai}$ is entirely negligible as compared with $\zeta_a |\mathbf{a}-\mathbf{b}|$, because their ratio is of order $\eta$ independently of any shape factor, while $\zeta_{ai}$ is exactly zero in the case of « axial plus planar » symmetry that applies very accurately to massive bodies of the solar system.) Now there remains the integral $K^{ai}$ [Eq. (A13)]. We note that, inside body (*a*), the $k_{ij}$ tensor [Eq. (2.31)] may be approximated as

$$k_{ij} = U h^{(a)}{}_{ij} [1 + O(\eta^2)], \qquad h^{(a)}{}_{ij} \equiv \frac{u_{a,i} u_{a,j}}{u_{a,k} u_{a,k}}, \tag{6.13}$$

because, inside ($a$), we have

$$\nabla U = \nabla u_a[1 + \mathrm{O}(\eta^2)]. \tag{6.14}$$

Under the current assumptions, the calculation of the integral $K^{ai}$ [eq. (A13)] does not involve any difficulty: the components of this integral are given by Eqs. (C10)-(C13), (C23). Taking benefit of these, and of Eqs. (6.10)-(6.12), we rewrite the equation for the PN correction to the motion of the mass centers, Eq. (4.9), as:

$$M_a^1 \ddot{a}_1^i = \alpha_a \frac{\partial U^{(a)}}{\partial a^i} + GM_a \frac{\partial}{\partial X^i}\bigg|_{\mathbf{X}=\mathbf{a}}\left(\sum_{b\neq a}\frac{\alpha_b}{|\mathbf{X}-\mathbf{b}|}\right) - (2/3)\varepsilon_a \ddot{a}^i$$

$$-\frac{GM_a}{2}\sum_{b\neq a}M_b\left(\ddot{b}^k\frac{\partial^2|\mathbf{a}-\mathbf{b}|}{\partial a^i\partial a^k}-\dot{b}^k\dot{b}^j\frac{\partial^3|\mathbf{a}-\mathbf{b}|}{\partial a^i\partial a^k\partial a^j}\right)-3\varepsilon_a U^{(a)}{}_{,i}(\mathbf{a})$$

$$+ (d/dT)\{-\dot{a}^i[2\varepsilon_a + M_a U^{(a)}(\mathbf{a})]/3 - [M_a\dot{\mathbf{a}}^2/2 + 2T_a + 4\varepsilon_a]\dot{a}^i - \gamma_a \Omega^{(a)}{}_{ji}\Omega^{(a)}{}_{jk}\dot{a}^k - M_a \dot{a}^i U^{(a)}(\mathbf{a})\}$$

$$+ [M_a U^{(a)}(\mathbf{a}) + 2\varepsilon_a]\Omega^{(a)}{}_{ji}\dot{a}^j - (\gamma_a/5)[\Omega^{(a)}{}_{mj}\Omega^{(a)}{}_{mj}U^{(a)}{}_{,i}(\mathbf{a}) + 2\Omega^{(a)}{}_{ij}\Omega^{(a)}{}_{lj}U^{(a)}{}_{,l}(\mathbf{a})]. \tag{6.15}$$

The PN correction to the mass is given by (A25), and the spherical symmetry of each body gives us

$$U^{(a)}(\mathbf{x}) = \Sigma_{b\neq a}\, GM_b/|\mathbf{x}-\mathbf{b}|. \tag{6.16}$$

Thus, Eq. (6.15) may be rewritten more explicitly as

$$M_a^1 \ddot{a}_1^i = G\frac{\partial}{\partial X^i}\bigg|_{\mathbf{X}=\mathbf{a}}\left(\sum_{b\neq a}\frac{\alpha_{ab}}{|\mathbf{X}-\mathbf{b}|}\right) - \frac{GM_a}{2}\sum_{b\neq a}M_b\left(\ddot{b}^k\frac{\partial^2|\mathbf{a}-\mathbf{b}|}{\partial a^i\partial a^k}-\dot{b}^k\dot{b}^j\frac{\partial^3|\mathbf{a}-\mathbf{b}|}{\partial a^i\partial a^k\partial a^j}\right)$$

$$- (d/dT)\left[\dot{a}^i\{(16/3)\varepsilon_a + 2T_a + M_a[(\dot{\mathbf{a}}^2/2)+(4/3)U^{(a)}(\mathbf{a})]\} + \gamma_a\Omega^{(a)}{}_{ji}\Omega^{(a)}{}_{jk}\dot{a}^k\right]$$

$$+ [M_a U^{(a)}(\mathbf{a}) + 2\varepsilon_a]\Omega^{(a)}{}_{ji}\dot{a}^j - (\gamma_a/5)[\Omega^{(a)}{}_{mj}\Omega^{(a)}{}_{mj}U^{(a)}{}_{,i}(\mathbf{a}) + 2\Omega^{(a)}{}_{ij}\Omega^{(a)}{}_{lj}U^{(a)}{}_{,l}(\mathbf{a})], \tag{6.17}$$

where

$$\alpha_{ab} \equiv M_a M_b\left(\frac{\dot{\mathbf{a}}^2+\dot{\mathbf{b}}^2}{2}+U^{(a)}(\mathbf{a})+U^{(b)}(\mathbf{b})\right) + M_b(M_a^1 + \frac{2}{3}\varepsilon_a + \frac{8}{3}T_a) + M_a(M_b^1 + \frac{11}{3}\varepsilon_b + \frac{8}{3}T_b). \tag{6.18}$$

When the (standard) PNA of GR is used for extended bodies, rotational terms do also occur, of course. Those include coupled terms like (6.6), that involve both translation and rotation velocities [5, Eqs. 78.04 and 78.06] — but these terms are neglected in practical calculations based on the Einstein-Infeld-Hoffmann equations [18]. As emphasized at the end of §6.1, this neglect does not *a priori* seem to be well-justified in the case of giant planets.

We shall not attempt here to further simplify this equation of motion, since anyway it must be entered in a computer program and this is clearly feasible: all quantities on the r.h.s. of (6.17) are known from the first-approximation calculation. Thus the PN corrections to the mass centers, $\mathbf{a}_1(T)$, shall be obtained just by (a double) *time integration.* However, we have to discuss the way in which the first-approximation quantities should be obtained.

## 7. Link with observational data. Remarks on the algorithm for numerical calculations

*7.1 General remarks on the link betwen observation and prediction in a theory of gravitation*

Let us consider some « relativistic » theory of gravitation, (T), for which one may define a precise Newtonian limit in terms of a parameter equivalent to the parameter $\lambda \equiv \varepsilon^2 \equiv (1 - f_{\min})/2$ in the scalar ether-theory [see Eq. (2.11); we have seen since that indeed the square $\varepsilon^2$ is more relevant, because the scalar theory admits expansions in powers of $\varepsilon^2$]. More precisely, let us assume that the equations and fields of theory (T) admit first-order expansions in terms of $\lambda$, with the zero-order equations reducing to NG, and with the remainder terms $O(\lambda^2)$ being negligible as compared with the observational error (note that this is currently the case as regards astrodynamics in the solar system, with $\varepsilon_0^4 \approx 10^{-12}$). Let $\mathbf{D} = (D_j)_{j=1,...,J}$ be a representative set of observational data (Keplerian parameters of planetary orbits and their time-drift, etc.). In order to perform the theoretical « prediction » of these data, theory (T) introduces some finite set of numerical parameters, $\boldsymbol{\alpha}$. Let $\boldsymbol{\alpha}^0 = (\alpha_i^0)_{i=1,...,I_0}$ be the subset which is needed for the zero-order calculation (this will include the first-approximation masses $M_a$ and some higher-order multipoles of the first-approximation mass density, etc.). Let $\boldsymbol{\alpha}^1 = (\alpha_i^1)_{i=1,...,I_1}$ be the complementary set which may be needed, in addition to $\boldsymbol{\alpha}^0$ (or rather to a part of $\boldsymbol{\alpha}^0$), for the first-order (second-approximation) correction. Thus

$$\boldsymbol{\alpha} = (\boldsymbol{\alpha}^0, \boldsymbol{\alpha}^1) \tag{7.1}$$

and the theoretical prediction for data $D_j$ is the sum of the zero-order (« Newtonian ») calculation and the first-order (PN) correction:

$$D_j^{\text{theory}} \equiv F_j(\boldsymbol{\alpha}) = F_j^0(\boldsymbol{\alpha}^0) + \lambda F_j^1(\boldsymbol{\alpha}). \tag{7.2}$$

(Of course, the functional expressions $F_j$ are not analytical ones and in practice will involve numerical calculations.)

Now it should be realized that the set of parameters $\boldsymbol{\alpha}$ can hardly be accessed independently of the set of observational data $\mathbf{D}$. For instance, it is clear that one cannot weigh the planets and that the masses $M_a$ are free parameters of the model [*i.e.,* of the first-order approximation of theory (T)]. Of course, some parameters such as the velocity of light (which does enter the expressions for PN corrections) *can* be accurately measured independently of any celestial mechanics, and we just remove them from the list $\boldsymbol{\alpha}$. Thus, the process of theoretical « prediction » involves essentially a *fitting of the observational data $D_j$ by the functional expressions $F_j$.* In practice, this will be a least-squares fitting, so one will have to search for the values of the free parameters $\boldsymbol{\alpha}$ that minimize the residual:

$$R(\boldsymbol{\alpha}) \equiv \Sigma_j [F_j(\boldsymbol{\alpha}) - D_j]^2 = \text{Min}. \tag{7.3}$$

But, in just the same way, what is called « Newtonian celestial mechanics » is essentially a fitting of the data $D_j$ by the first-approximation expressions $F_j^{\,0}$, and this gives the *Newtonian values* $\boldsymbol{\alpha}^N = (\alpha_i^N)_{i=1,\ldots,I_0}$ *of the zero-order parameters:*

$$R^0(\boldsymbol{\alpha}^N) \equiv \Sigma_j\, [F_j^{\,0}(\boldsymbol{\alpha}^N) - D_j]^2 = \text{Min.} \qquad (7.4)$$

The Newtonian values of the zero-order parameters have no reason to be optimal for the PN calculation, simply because the functional expressions to be minimized are different. In particular, the Newtonian masses $M_a^N$ have no reason to give adequate values of the zero-order masses $M_a$ for PN calculations.

To get some idea about the difference that makes, it is reasonable to assume: **(i)** that *« the theory (T) is good »* in the sense that its PN approximation (with the optimal set of parameters $\boldsymbol{\alpha}$) gives predictions for **D** which coincide with **D** up to an $O(\lambda^2)$ error [thus, the improvement of NG by theory (T) should not be found merely in the solar system, but instead should be verified each time one tests a weakly gravitating system]; **(ii)** that *« NG is not bad »* in the sense that the difference between the Newtonian values $\alpha_i^N$ of the zero-order parameters (values that are optimal for the first approximation) and their correct values $\alpha_i^0$ (that are optimal for the second approximation) is *small*; and **(iii)** that, however, *« the theory (T) is necessary »* in the sense that the difference between the Newtonian prediction $F_j^{\,0}(\boldsymbol{\alpha}^N)$ and the observed data $D_j$ is ord($\lambda$). {We adopt the varying units of mass and time, $[M]_\varepsilon \equiv \varepsilon^2[M]$ and $[T]_\varepsilon \equiv \varepsilon^{-1}[T]$ (§2.3), so that **D**, $\boldsymbol{\alpha}$, $\mathbf{F}^0(\boldsymbol{\alpha}^0)$ and $\mathbf{F}^1(\boldsymbol{\alpha})$ are ord($\lambda^0$).} From (i) and (ii), we get by (7.2):

$$F_j^{\,0}(\boldsymbol{\alpha}^N) + (\partial F_j^{\,0}/\partial \alpha_i^N)\, \delta\alpha_i^0 + \lambda F_j^{\,1}(\boldsymbol{\alpha}) = D_j + O(\lambda^2), \qquad \delta\alpha_i^0 \equiv \alpha_i^0 - \alpha_i^N. \qquad (7.5)$$

Hence, by (iii), the quantities

$$(\partial F_j^{\,0}/\partial \alpha_i^N)\, \delta\alpha_i^0 = [D_j - F_j^{\,0}(\boldsymbol{\alpha}^N)] - \lambda F_j^{\,1}(\boldsymbol{\alpha}) + O(\lambda^2) \qquad (7.6)$$

are $O(\lambda)$ anyway, and should *a priori* be ord($\lambda$), unless miraculously the $\boldsymbol{\alpha}^N$ were systematically optimal zero-order parameters for the first-order calculation. Since $\partial F_j^{\,0}/\partial \alpha_i^N = \text{ord}(\lambda^0)$ in the varying units, this means that, in the same way, $\delta\alpha_i^0$ should be ord($\lambda$) [and is $O(\lambda)$ anyway]. If, in the PN fitting (7.3), one assigns the Newtonian optimal values $\boldsymbol{\alpha}^N$ to the zero-order parameters, one shall get values $\boldsymbol{\alpha}'^1 \neq \boldsymbol{\alpha}^1$ and so the whole set of PN parameters shall be $\boldsymbol{\alpha}' \equiv (\boldsymbol{\alpha}^N, \boldsymbol{\alpha}'^1) \neq \boldsymbol{\alpha}$. Then, the total PN prediction obtained so will differ from the optimal PN prediction by:

$$\delta D_j \equiv F_j^{\,0}(\boldsymbol{\alpha}^N) + \lambda F_j^{\,1}(\boldsymbol{\alpha}') - [F_j^{\,0}(\boldsymbol{\alpha}^0) + \lambda F_j^{\,1}(\boldsymbol{\alpha})] = (\partial F_j^{\,0}/\partial \alpha_i^N)\, \delta\alpha_i^0 + \lambda[F_j^{\,1}(\boldsymbol{\alpha}') - F_j^{\,1}(\boldsymbol{\alpha})] \quad (7.7)$$

and so, *a priori,* will be ord($\lambda$). Thus, *the error made in taking the Newtonian optimal values $\boldsymbol{\alpha}^N$ for the zero-order parameters is* a priori *of the same order* [ord($\lambda$)] *as the PN correction*. Admittedly, we cannot exclude the case that a « miraculous » cancellation would make the difference (7.7) $O(\lambda^2)$, hence acceptable for the first PNA, but in the absence of any proof of that, we cannot assign the Newtonian optimal values $\boldsymbol{\alpha}^N$ to the zero-order parameters $\boldsymbol{\alpha}^0$. This applies to any theory of gravity admitting asymptotic expansions for weak fields and whose zero-order approximation reduces to NG.

*7.2 Relevant parameters in the scalar theory and the question of the preferred frame*

Let us list the parameters needed to perform the calculation of the PN corrections as given by Eq. (6.17). These are the first-approximation masses $M_a$, and other integrals of the first-approximation fields: $T_a$ [Eq. (5.10)], $\varepsilon_a$ [Eq. (5.12)], $\gamma_a$ [Eq. (6.5)] and $M_a^1$ [Eq. (A25)]. Note that the presence of these integrals means unambiguously that the internal structure of the bodies influences the motion at the level of the PN corrections. Moreover, in (6.17), we have *functions* of the absolute time $T$ (*i.e.* the inertial time in the preferred reference frame E), namely the first-approximation positions, velocities and accelerations of the bodies, $\mathbf{a}(T)$, $\dot{\mathbf{a}}(T)$ and $\ddot{\mathbf{a}}(T)$ and the rotation velocity tensors $\mathbf{\Omega}^{(a)}$ for $a = 1, ..., N$ (actually, the latter tensors may be considered constant for the purpose of PN corrections at the accuracy envisaged here). But the accelerations are given by Eq. (4.6), hence are determined by the masses $M_a$, the inertia tensors $\mathbf{I}^{(a)}$, [and if desired some higher-order multipoles $\mathbf{I}_{higher}^{(a)}$ of the zero-order mass density $\rho_a$,] plus the current positions $\mathbf{a}(T)$. To determine the zero-order problem, we also need the initial velocities and positions. Thus, the list of the zero-order parameters is as follows:

$$\alpha^0 = \left(M_a, \mathbf{I}^{(a)}{}_0, [\mathbf{I}_{higher}^{(a)}{}_0,] \mathbf{a}_0, \dot{\mathbf{a}}_0, \mathbf{\Omega}^{(a)}{}_0 \quad (a = 1, ..., N)\right), \tag{7.8}$$

where index 0 means initial value. (Since the bodies are assumed to have a rigid motion, the inertia tensor $\mathbf{I}^{(a)}$ [and $\mathbf{I}_{higher}^{(a)}$] merely rotates with the rotation velocity $\mathbf{\Omega}^{(a)}$.)

The positions, velocities and accelerations have to be given in the frame E, yet the astronomical observations, *e.g.* in the solar system, are expressed in some moving frame F – such as Copernic's reference frame C, whose axes are defined with respect to the « fixed stars », and whose origin is at the global mass center of the solar system. In order to properly define the correspondence between observables in the different frames involved, say E and F, we shall assume that F undergoes a *pure translation*, at some velocity $\mathbf{V}(T)$, with respect to the preferred frame E. This means that, as seen from E (and, in particular, using the « absolute simultaneity », defined by the Poincaré-Einstein synchronization applied in the frame E), each observer in F has the same velocity $\mathbf{V}(T)$. For observations in the solar system, the relevant absolute velocity $\mathbf{V}$ shall be that of its mass center, *i.e.* one will have F = C. For PN corrections in the solar system, the relevant time scale is the century [for larger time intervals, one should « reinitialize » the problem, see around Eq. (3.14)]. Therefore, we shall assume here that $\mathbf{V}$ *is a constant,* so that the correspondence between E and F is a Lorentz transformation of the flat metric $\gamma^0$ [14] – a special one, if one takes the $Ox^1$ and $O'x'^1$ axes parallel to $\mathbf{V}$:

$$x'^1 = \gamma_V (x^1 - VT), \qquad T' = \gamma_V (T - Vx^1/c^2), \quad x'^2 = x^2, \qquad x'^3 = x^3. \tag{7.9}$$

Hence, relevant physical fields in the moving frame (now denoted $E_V$) are the transported *proper fields:*

$$p'_{exact}(T', \mathbf{x}') \equiv p_{exact}[T(T', \mathbf{x}'), \mathbf{x}(T', \mathbf{x}')] \tag{7.10}$$

and the like for $\rho^*$ and $\Pi$, plus the *Lorentz-transformed* velocity and acceleration: $\mathbf{u}'_{exact} \equiv d\mathbf{x}'/dT'$ and $\mathbf{A}'_{exact} \equiv d\mathbf{u}'_{exact}/dT'$. As to the gravitational field, we shall consider the transported field

$$f'(T', \mathbf{x}') \equiv f[T(T', \mathbf{x}'), \mathbf{x}(T', \mathbf{x}')], \tag{7.11}$$

though it has not a direct physical significance. Since the fields in the frame E admit asymptotic expansions, we may obtain ones for the fields in the frame $E_V$. Because we must assume that $V/c = O(\varepsilon)$ [14], it follows that the Lorentz transformation (7.9) involves the small parameter. Therefore, we have to account for the expansion of (7.9). Neglecting $O(\varepsilon^4)$, the latter is:

$$x'^1 = [1 + V^2/(2c^2)](x^1 - VT) [1 + O(\varepsilon^4)], \qquad T' = \{T + (V^2/c^2)[(T/2) - (x^1/V)]\}[1 + O(\varepsilon^4)], \qquad (7.12)$$

where it has been used the fact that $\mathbf{x}$ is in the near zone, i.e. $\mathbf{x} = O(\varepsilon^0)$. In particular, when one neglects $O(\varepsilon^2)$, it becomes the Galileo transformation:

$$\mathbf{x}' = (\mathbf{x} - \mathbf{V}T)[1 + O(\varepsilon^2)], \quad T' = T[1 + O(\varepsilon^2)]. \qquad (7.13)$$

We now have to insert the initial expansions into the expressions of the transported fields, e.g. insert (2.17) into (7.10), and to account for (7.12) and (7.13). This provides an expansion for the transported fields: e.g.

$$p'_{exact}(T', \mathbf{x}') \equiv p_{exact}[T(T', \mathbf{x}'), \mathbf{x}(T', \mathbf{x}')] = [p(T, \mathbf{x}) + p_1(T, \mathbf{x})/c^2][1 + O(\varepsilon^4)]$$

$$= [p'(T', \mathbf{x}') + p'_1(T', \mathbf{x}')/c^2][1 + O(\varepsilon^4)]. \qquad (7.14)$$

Due to (7.13), the zero-order term of the transported field is simply the Galileo transform of the initial zero-order term, i.e.

$$p'(T', \mathbf{x}') = p(T', \mathbf{x}' + \mathbf{V}T') \qquad (7.15)$$

so that the notation $p'$ is unambiguous. The same is true for $\rho^*$ and $\Pi$, of course. In the same way, we have $U'(T', \mathbf{x}') = U(T', \mathbf{x}' + \mathbf{V}T')$ since, from (7.11) and (7.13):

$$f'(T', \mathbf{x}') \equiv f(T, \mathbf{x}) = 1 - 2U(T, \mathbf{x})/c^2 + O(\varepsilon^4) = 1 - 2U(T', \mathbf{x}' + \mathbf{V}T')/c^2 + O(\varepsilon^4). \qquad (7.16)$$

Moreover, it is easy to show [14] that the *Lorentz-transformed* components $\gamma'_{\mu\nu}$ of the physical metric $\gamma$ satisfy

$$\gamma'_{00}(T', \mathbf{x}') = 1 - 2U'(T', \mathbf{x}')/c^2 + O(\varepsilon^4), \qquad (7.17)$$

so that the transported potential $U'$ has a direct physical significance. As to the velocity and acceleration, they were not defined by a mere transport, but we get easily for the zero-order terms:

$$\mathbf{u}'(T', \mathbf{x}') = \mathbf{u}(T', \mathbf{x}' + \mathbf{V}T') - \mathbf{V}, \qquad \mathbf{A}'(T', \mathbf{x}') = d\mathbf{u}'/dT' = (d\mathbf{u}/dT)(T', \mathbf{x}' + \mathbf{V}T'). \qquad (7.18)$$

The equations for the exact fields in the moving frame $E_V$ are obtained by substituting for the fields in the preferred frame E their expressions in terms of the fields in $E_V$, like (7.10) and (7.11), in Eqs. (2.3), (2.6) and (2.7). Using the Lorentz transform, one expresses the derivatives in E as derivatives in $E_V$ [14, Eq (49)]. As one may easily check, it follows from these definitions that *the zero-order fields in the frame $E_V$ obey the equations of NG,* Eqs. (2.25), (2.27) and (2.29) with primes. This was expected since (i) the zero-order fields in the frame E obey NG, (ii) the zero-order fields in E and in $E_V$ exchange by Galileo transformation, and (iii) NG is Galileo-invariant.

*7.3 The gross structure of an algorithm for celestial mechanics in the scalar theory*

Having found a firm basis for NG in the scalar theory, we may use it, in particular, to assess the integrals $T_a$, $\varepsilon_a$, $\gamma_a$ and $M_a^1$ [Eqs. (5.10), (5.12), (6.5) and (A25)]. Thus, in the solar system, we may use Newtonian estimates of the density profile $\rho = \varphi_a(r)$ for the different bodies (*a*) [30-31] and compute these integrals. As emphasized in §7.1, such purely Newtonian calculations of the first-approximation quantities are correct only up to an $O(\varepsilon^2)$ error, but this does little matter here, because $\varepsilon_a$, $\xi_a$ and $\theta_a$ are used only in second-approximation (PN) *corrections*, hence an $O(\varepsilon^2)$ error on them makes an $O(\varepsilon^4)$ (third-approximation) error in the total PN estimate. Thus, *the PN parameters $\alpha^1$ reduce to the absolute velocity of the moving frame* $E_V$ *used for observations,* i.e. *to the vector* **V**. In practice, that moving frame could be Copernic's frame C. However, most observations are performed, not in a such uniformly moving frame directly, but instead in local frames with a more complex motion, *e.g.* with Earth-based telescopes. In GR, formalisms exist to pass from « local » frames to « global » ones [32, 33]. We shall not discuss here the way this correspondence has to be modified in the scalar theory and will assume that the observables **D** are expressed in some uniformly moving (« inertial ») frame $E_V$ − but **V** is not *a priori* known.

The skeleton of an algorithm may then be proposed for PN calculations in the scalar theory:

0) Initialization: Perform the purely Newtonian calculation, *i.e.* solve (7.4) by iterating the (non-linear) zero-order calculation alone, and thus obtain the Newtonian values $\alpha^N$ of the zero-order parameters (7.8). Of course, it will be more convenient to make that step in $E_V$. Since the list (7.8) is given in E, one will have to Galileo-transform the zero-order parameters from $E_V$ to E. As it appears from (7.8), only the velocities $\dot{\mathbf{a}}_0$ shall be affected: they shall be modified according to (7.18)$_1$.

1) a) Some estimate for the zero-order parameters $\alpha^0$ and for **V** being known, compute the zero-order motion (as was done for each iteration of step 0), and compute the PN corrections to the motion by (6.17). Then Lorentz-transform the total motion, to $O(1/c^2)$ accuracy, back to $E_V$. Estimate the residual (7.3).
   b) Using a minimization algorithm, make loops on a) to optimize $\alpha^0$ and **V**.

Two different starting estimates seem possible for **V**: either **0**, or the velocity against the cosmic microwave background. The decisive points for the theory shall be (i) the accuracy, *i.e.* the level of the final value of the residual (7.3); *and* (ii) the magnitude of the optimal velocity vector **V**. Since the theory predicts exact Schwarzschild motion in the spherical static case, one is tempted to conjecture that a correct accuracy (improving significantly NG) should be reached. But, for which vector **V**? If $V = |\mathbf{V}|$ were found too small (some km/s), that would mean Copernic's reference frame is nearly at rest in the ether − which seems difficult to justify. On the other hand, if a good agreement were found for a significant value of *V*, that would be a strong argument for an ether.

# Appendix A. Integrals $I^{ai}$, $J^{ai}$, $K^{ai}$ and $L^{ai}$ and $M_a^1$ (well-separated rigid bodies)

To evaluate integrals which are similar to $I^{ai}$ and $J^{ai}$, Fock [5, §§ 76-77] uses the relations

$$\rho \Pi - \rho u_a + p = \rho \Omega_a, \tag{A1}$$

$$\mathbf{u}^2/2 = \dot{\mathbf{a}}^2/2 + \dot{a}^k \Omega^{(a)}{}_{jk}(x^j - a^j) + \Omega_a, \tag{A2}$$

valid inside body (*a*). Equation (A2) follows from the particular form (5.4) assumed for the velocity field, and Eq. (A1) is deduced from the Newtonian equation of motion (2.29), accounting for (5.4) and for the expression of the elastic energy in a barotropic fluid:

$$\Pi(p) = \int_0^p \frac{\mathrm{d}q}{\rho^*(q)} - \frac{p}{\rho^*(p)}. \tag{A3}$$

Let us evaluate $J^{ai}$. It follows from (A1) and (A2) that

$$\rho (\mathbf{u}^2/2 + \Pi + u_a) = \rho [\dot{\mathbf{a}}^2/2 + \dot{a}^k \Omega^{(a)}{}_{jk}(x^j - a^j) + 2\Omega_a + 2 u_a] - p. \tag{A4}$$

Using the definitions (5.10)-(5.14) and Fock's integrals (74.24) and (74.25):

$$3 \int_{\omega_a} p \, \mathrm{d}V = \varepsilon_a - 2 T_a, \tag{A5}$$

$$2 \int_{\omega_a} p (x^i - a^i) \, \mathrm{d}V = \eta_{ai} - T_{ai}, \tag{A6}$$

we get then [remembering the definition (3.9) of the mass center]

$$\int_{\omega_a} \rho (\mathbf{u}^2/2 + \Pi + u_a) \, \mathrm{d}V = M_a \dot{\mathbf{a}}^2/2 + (8T_a + 11\varepsilon_a)/3 \tag{A7}$$

and

$$\int_{\omega_a} \rho (\mathbf{u}^2/2 + \Pi + u_a)(x^i - a^i) \, \mathrm{d}V = \dot{a}^k \Omega^{(a)}{}_{jk} I^{(a)}{}_{ij} + (5T_{ai} + 8\varepsilon_{ai} - \eta_{ai})/2. \tag{A8}$$

We have from (5.1):

$$\int_{\omega_b} \varphi /|\mathbf{X} - \mathbf{x}| \, \mathrm{d}V(\mathbf{x}) = [\int_{\omega_b} \varphi \, \mathrm{d}V/|\mathbf{X} - \mathbf{b}| + (X^i - b^i) \int_{\omega_b} \varphi(\mathbf{x})(x^i - b^i) \mathrm{d}V(\mathbf{x}) /|\mathbf{X} - \mathbf{b}|^3][1+ O(\eta^2)]. \tag{A9}$$

We apply this to $\varphi \equiv \varphi_b + \rho U^{(b)}$ and use (5.3). For $\varphi_b$, we use (A7)-(A8) and (3.9)$_2$; for $\rho U^{(b)}$, we make a Taylor expansion of $U^{(b)}$ at **b** and use (3.9)$_1$. This gives us Eq. (5.5). Making a Taylor expansion of $B^{(a)}$ at **a** and using (3.9)$_1$, we get (5.6). Equation (5.7) is obtained in the same way.

To calculate $I^{ai}$ [Eq. (4.8)], we substitute the expression (5.4) for **u** in (4.8) and use (A5)-(A8) to obtain

$$\int_{\omega_a} [p + \rho (\mathbf{u}^2/2 + \Pi + u_a)] u^i \, \mathrm{d}V = [M_a \dot{\mathbf{a}}^2/2 + 2T_a + 4\varepsilon_a]\dot{a}^i + (\dot{a}^k \Omega^{(a)}{}_{lk} I^{(a)}{}_{jl} + 2T_{aj} + 4\varepsilon_{aj})\Omega^{(a)}{}_{ji}. \tag{A10}$$

Making a Taylor expansion of $U^{(a)}$ at **a** and using (3.9)$_1$, we get

$$\int_{\omega_a} \rho\, U^{(a)}\, u^i\, dV = [M_a\, \dot{a}^i\, U^{(a)}(\mathbf{a}) + I^{(a)}{}_{jk}\, \Omega^{(a)}{}_{ki}\, U^{(a)}{}_{,j}(\mathbf{a})]\, [1 + O(\eta^2)], \tag{A11}$$

and $I^{ai}$ is obtained by summing (A10) and (A11).

As to $K^{ai}$: using $(2.31)_1$ and applying (4.1) with $z^i \equiv \rho k_{ij} u^j$, neglecting $\int_{\partial\omega_a} z^i\, \mathbf{u}_1.\mathbf{n}\, dS/c^2$ (as is plainly justified, see Section 3), we find by (2.29) that

$$\int_{\omega_a} ({}_1\Gamma^i_{jk}\, \rho u^j\, u^k + \rho u^j \partial_T k_{ij})\, dV = \int_{\omega_a} \rho u^j [(\partial_T k_{ij} + k_{ij,k}\, u^k) + (k_{ik,j} - k_{jk,i})\, u^k] dV$$
$$= (d/dT)\left(\int_{\omega_a} \rho\, k_{ij} u^j\, dV\right) - \int_{\omega_a} k_{ij}\, (\rho U_{,j} - p_{,j})\, dV + \int_{\omega_a} \rho u^j (k_{ik,j} - k_{jk,i})\, u^k\, dV. \tag{A12}$$

Substituting this in the definition of $K^{ai}$ (4.10), we get:

$$K^{ai} = \int_{\omega_a} k_{ij}(p_{,j} + \rho U_{,j}) dV + \int_{\omega_a} p U_{,i} dV + \int_{\omega_a} (-2U\rho U_{,i}) dV - (d/dT)\left(\int_{\omega_a} \rho k_{ij} u^j dV\right) + \int_{\omega_a} \rho u^j u^k (k_{jk,i} - k_{ik,j}) dV$$
$$= K_1^{ai} + K_2^{ai} + K_3^{ai} - (d/dT)K'^{ai} + K_4^{ai}. \tag{A13}$$

We decompose $L^{ai}$ into « self » and « external » parts:

$$L^{ai} = \int_{\omega_a} \rho\, (\partial^3 w_a / \partial x^i \partial T^2)\, dV + \int_{\omega_a} \rho\, (\partial^3 W^{(a)} / \partial x^i \partial T^2)\, dV. \tag{A14}$$

The transformation of the « self » part is rather involved. From (4.16), it follows that

$$\int_{\omega_a} \tau u^*{}_{a,i}\, dV = \int_{\omega_a} \tau\, U^{\dagger(a)}{}_{,i}\, dV + \left[\int_{\omega_a} \rho\, (\partial^3 W^{(a)}/\partial x^i \partial T^2)\, dV + \int_{\omega_a} \rho\, (\partial^3 w_a/\partial x^i \partial T^2)\, dV\right]/c^2$$
$$- \int_{\omega_a} \tau\, U^{*(a)}{}_{,i}\, dV + O(\varepsilon^8), \tag{A15}$$

and since $U^{*(a)} = U^{\dagger(a)} + (\partial^2 W^{(a)}/\partial T^2)/c^2$, we get

$$\int_{\omega_a} \rho\, (\partial^3 w_a/\partial x^i \partial T^2)\, dV/c^2 = \int_{\omega_a} \tau u^*{}_{a,i}\, dV + O(\varepsilon^8). \tag{A16}$$

It is proved in Appendix B that

$$\int_{\omega_a} \tau u^*{}_{a,i}\, dV = \int_{\text{space}} u^*{}_{a,i}\, (\partial^2 u_a/\partial T^2)\, dV\,]\, /(4\pi G c^2), \tag{A17}$$

$$\int_{\text{space}} u^*{}_{a,i}\, (\partial^2 u_a/\partial T^2)\, dV = \int_{\text{space}} u_{a,i}\, (\partial^2 u_a/\partial T^2)\, dV + O(\varepsilon^8), \tag{A18}$$

$$\int_{\text{space}} u_{a,i}\, (\partial^2 u_a/\partial T^2)\, dV = (d/dT) \int_{\omega_a} 4\pi G \rho\, (\partial^2 w_a/\partial x^i \partial T)\, dV. \tag{A19}$$

From (A16)-(A19), we obtain

$$\int_{\omega_a} \rho\, (\partial^3 w_a/\partial x^i \partial T^2)\, dV = (d/dT) \int_{\omega_a} \rho\, (\partial^2 w_a/\partial x^i \partial T)\, dV + O(\varepsilon^8), \tag{A20}$$

which we were not able to prove in a simpler way. The integral on the r.h.s. is calculated by Fock [5, Eq. (76.11)], which gives

$$\int_{\omega_a} \rho\,(\partial^3 w_a/\partial x^i\,\partial T^2)\,dV = (d/dT)[-\varepsilon_a\,\dot{a}^i + B^{(a)}{}_{ik}\,\dot{a}^k - \Omega^{(a)}{}_{ji}\varepsilon_{aj} + \Omega^{(a)}{}_{jk}\,B^{(a)}{}_{j\,ik}]. \quad (A21)$$

As to the « external » part of $L^{ai}$, we get by a Taylor expansion and $(3.9)_1$:

$$\int_{\omega_a} \rho\,(\partial^3 W^{(a)}/\partial x^i\,\partial T^2)\,dV = [M_a\,(\partial^3 W^{(a)}/\partial x^i\,\partial T^2)(\mathbf{a})]\,[1+ O(\eta^2)]. \quad (A22)$$

Using Fock's formula (76.34):

$$\frac{\partial^2 W^{(a)}}{\partial x^i\,\partial T} = -\frac{1}{2}\sum_{b\neq a} GM_b \dot{b}^k \frac{\partial^2 |\mathbf{x}-\mathbf{b}|}{\partial x^i\,\partial x^k} + \frac{1}{4}\sum_{b\neq a} G\dot{I}^{(b)}_{jk}\frac{\partial^3 |\mathbf{x}-\mathbf{b}|}{\partial x^i\,\partial x^j\,\partial x^k}, \quad (A23)$$

we obtain thus [neglecting $O(\eta^2)$ as compared with $O(\eta^0)$, as also does Fock to get (A23)]:

$$\int_{\omega_a} \rho\frac{\partial^3 W^{(a)}}{\partial x^i\,\partial T^2}\,dV = -\frac{GM_a}{2}\left[\sum_{b\neq a} M_b\left(\ddot{b}^k\frac{\partial^2 |\mathbf{a}-\mathbf{b}|}{\partial a^i\,\partial a^k} - \dot{b}^k\dot{b}^j\frac{\partial^3 |\mathbf{a}-\mathbf{b}|}{\partial a^i\,\partial a^k\,\partial a^j}\right)\right.$$
$$\left. -\frac{1}{2}\sum_{b\neq a}\left(\ddot{I}^{(b)}_{jk}\frac{\partial^3 |\mathbf{a}-\mathbf{b}|}{\partial a^i\,\partial a^j\,\partial a^k} - \dot{I}^{(b)}_{jk}\dot{b}^l\frac{\partial^4 |\mathbf{a}-\mathbf{b}|}{\partial a^i\,\partial a^j\,\partial a^k\,\partial a^l}\right)\right]. \quad (A24)$$

According to (A14), $L^{ai}$ is obtained by summing (A21) and (A24).

Finally, we get by inserting (A2) into (3.16) and with the usual Taylor expansion of $U^{(a)}$:

$$M_a^1 = [\{M_a[\dot{\mathbf{a}}^2/2 + U^{(a)}(\mathbf{a})] + T_a\}_{T=0} + 2\varepsilon_a]\,[1+ O(\eta^2)]. \quad (A25)$$

Note that $\varepsilon_a$ is a constant as far as we assume that $(a)$ has a rigid motion.

### Appendix B. Proof of the equalities (A17), (A18) and (A19)

*Proof of Eq. (A17).* Since $U^* \equiv U + A/c^2$, we have by (2.25) and (2.26):

$$-4\pi G\tau = \Delta U^* - \partial^2 U/\partial T^2/c^2, \quad (B1)$$

whence [remembering the definition (4.13) for the self fields]

$$4\pi G \int_{\omega_a} \tau\,u^*_{a,i}\,dV = \int_{\text{space}} u^*_{a,i}\,(\partial^2 u_a/\partial T^2)\,dV/c^2 - \int_{\text{space}} u^*_{a,i}\,\Delta u^*_a\,dV. \quad (B2)$$

In this work, we define the integral over the whole space as the limit (when it exists):

$$\int_{\text{space}} \theta\,dV \equiv \lim_{r\to\infty} \int_{|\mathbf{x}-\mathbf{A}|\leq r} \theta\,dV, \quad (B3)$$

which depends, in general, on the choice of the origin **A**. For the integrals considered, we shall take **A** = **a**, *i.e.* the first-approximation mass center of body (*a*), but actually these integrals do not depend on the origin. We obtain thus by the divergence theorem:

$$\int_{\text{space}} u^*_{a,i} \Delta u^*_a dV = \lim_{r \to \infty} \int_{S_r} [n_j u^*_{a,i} u^*_{a,j} - n_i u^*_{a,k} u^*_{a,k}/2] dS = \lim_{r \to \infty} \int_{S_r} [(\mathbf{g}^* \cdot \mathbf{n}) \mathbf{g}^* - \mathbf{g}^{*2} \mathbf{n}/2] dS \quad (B4)$$

with $\mathbf{g}^* \equiv \nabla u^*_a$ and where $S_r$ is the sphere $|\mathbf{x} - \mathbf{a}| = r$. Since $u^*_a = \partial^2 w_a/\partial T^2/c^2 + \text{N.P.}[\tau_a]$, we have $\mathbf{g}^* \equiv \nabla(\partial^2 w_a/\partial T^2/c^2) + O(1/r^2)$. Now by expanding $|\mathbf{x} - \mathbf{y}|$ in a Taylor series at $\mathbf{x} - \mathbf{a}$, we get:

$$w_a(\mathbf{x}, T) \equiv \int_{\omega_a} G|\mathbf{x} - \mathbf{y}|\rho(\mathbf{y},T) dV(\mathbf{y})/2 = GM_a|\mathbf{x} - \mathbf{a}|/2 + GI^{(a)}_{ij}(\delta_{ij} - n_i n_j)/(4|\mathbf{x} - \mathbf{a}|) + ..., \quad (B5)$$

$$\partial_T w_a = -GM_a \mathbf{n} \cdot \dot{\mathbf{a}}/2 + O(1/r), \qquad \partial_T^2 w_a = -GM_a \mathbf{n} \cdot \ddot{\mathbf{a}}/2 + O(1/r), \quad (B6)$$

$$\nabla(\partial^2 w_a/\partial T^2) = -GM_a \mathbf{z}(\mathbf{n})/(2r) + O(1/r^2). \quad [\,\mathbf{z}(\mathbf{n}) \equiv \ddot{\mathbf{a}} - (\mathbf{n} \cdot \ddot{\mathbf{a}})\mathbf{n}, \quad \mathbf{n} \equiv (\mathbf{x} - \mathbf{a})/r, \quad r \equiv |\mathbf{x} - \mathbf{a}|.\,] \quad (B7)$$

Since $\mathbf{z} \cdot \mathbf{n} = 0$ and since $\mathbf{z}(\mathbf{n}) = \mathbf{z}(-\mathbf{n})$, it follows that the limit in (B4) is zero. This proves (A17).

*Proof of Eq. (A18).* As for $\partial_T^2 w_a$, one finds that

$$u_a(\mathbf{x}, T) \equiv \int_{\omega_a} G\rho(\mathbf{y},T) dV(\mathbf{y})/|\mathbf{x} - \mathbf{y}| = GM_a/|\mathbf{x} - \mathbf{a}| + GI^{(a)}_{ij}(3n_i n_j - \delta_{ij})/(2|\mathbf{x} - \mathbf{a}|^3) + ..., \quad (B8)$$

$$\partial_T u_a = GM_a \mathbf{n} \cdot \dot{\mathbf{a}}/r^2 + O(1/r^3), \qquad \partial_T^2 u_a = GM_a \mathbf{n} \cdot \ddot{\mathbf{a}}/r^2 + O(1/r^3) \quad (r \to \infty). \quad (B9)$$

At large $r$, $u_{a,i}$ is ord($1/r^2$) and, as we have seen, $u^*_{a,i}$ is ord($1/r$). Moreover, both integrands in (A18) are bounded. Thus, the integrand on the r.h.s. of (A18) is (Lebesgue-)integrable, being $O(1/r^4)$ at large $r$, so that the integral on the r.h.s. *a fortiori* exists in the sense of (B3). The integrand on the l.h.s. is ord($1/r^3$), hence not integrable. However, the main term in that integrand is

$$\text{Const} \times [\mathbf{w} - (\mathbf{w} \cdot \mathbf{n})\mathbf{n}] \, \mathbf{w} \cdot \mathbf{n}/r^3, \qquad [\,\mathbf{w} \equiv \ddot{\mathbf{a}}\,] \quad (B10)$$

hence changes sign with **n** and so makes a zero net contribution in each ball $|\mathbf{x} - \mathbf{a}| \leq r$, thus admitting zero as the limit. Since the remainder is $O(1/r^4)$, therefore integrable, it follows that the integral on the l.h.s of (A18) also exists in the sense of (B3). Hence, the difference between the two terms:

$$\int_{\text{space}} (u^*_{a,i} - u_{a,i})(\partial^2 u_a/\partial T^2) \, dV = \int_{\text{space}} A_{a,i} (\partial^2 u_a/\partial T^2) \, dV/c^2 \equiv D_{ai} \quad (B11)$$

also exists in the sense of (B3). In the varying units introduced in §2.3, $D_{ai}$ is obviously $O(\varepsilon^2)$, and since $D_{ai}$ has dimension $L^6 T^{-6}$, it is therefore $O(\varepsilon^8)$ in the fixed units. This proves (A18).

*Proof of Eq. (A19).* We have by the divergence theorem:

$$\int_{\text{space}} u_{a,i} \partial_T^2 u_a \, dV = \int_{\text{space}} [\partial_T(\partial_i u_a \partial_T u_a) - \partial_T u_a \partial_T \partial_i u_a] \, dV$$

$$= (d/dT) \int_{\text{space}} \partial_i u_a \partial_T u_a \, dV - \lim_{r \to \infty} \int_{S_r} [\partial_T (u_a^2)/2] \, n_i \, dS \,. \quad (B12)$$

We used an inversion between integration and time differentiation. In order to justify this by Lebesgue's theorem, we must show that: i) for any $T$, $\partial_i u_a \partial_T u_a$ is integrable, and ii) $\partial_T(\partial_i u_a \partial_T u_a)$ is bounded, independently of $T$, by an integrable function $\phi(\mathbf{x})$. Condition i) is satisfied, for $\partial_i u_a \partial_T u_a$ is bounded, and $O(1/r^4)$ at large $r$. As regards condition ii), we note that, due to the assumed quasi-periodic character of the motion (in some interval of time I), $\partial_T (\partial_i u_a \partial_T u_a)$ is bounded, independently of $\mathbf{x}$ and $T$, by some number $\alpha$. Moreover, for any $T$, it has the form [*cf.* (B8) and (B9)]

$$\partial_T(\partial_i u_a \partial_T u_a)(\mathbf{x},T) = -G^2 M_a^2 n_i n_j \ddot{a}^j / |\mathbf{x} - \mathbf{a}|^4 + R(T, \mathbf{x}) \quad (B13)$$

with

$$R(T, \mathbf{x}) = O(1/r^5) \quad \text{as } r \equiv |\mathbf{x} - \mathbf{a}| \to \infty. \quad (B14)$$

Again due to the assumed quasi-periodic character of the motion, the behaviour of the remainder term is valid independently of $T$; *i.e.*, there are some numbers $M$ and $r_0$ such that

$$\forall T \in I \qquad R(T, \mathbf{x}) \leq M/|\mathbf{x} - \mathbf{a}|^5 \quad \text{if} \quad |\mathbf{x} - \mathbf{a}| > r_0. \quad (B15)$$

Hence $\partial_T(\partial_i u_a \partial_T u_a)(\mathbf{x},T)$ is bounded, for $|\mathbf{x} - \mathbf{a}| > r_0$, by the function

$$\phi_i(\mathbf{x}) \equiv M/|\mathbf{x} - \mathbf{a}|^5 + G^2 M_a^2 |n_i n_j| \{\text{Sup}_{T \in I}[|\ddot{a}^j(T)|]\}/|\mathbf{x} - \mathbf{a}|^4 \quad (B16)$$

which is integrable on the domain $|\mathbf{x} - \mathbf{a}| > r_0$. Thus, defining $\phi_i(\mathbf{x}) \equiv \alpha$ for $|\mathbf{x} - \mathbf{a}| \leq r_0$, one satisfies condition (ii), so the inversion between integration and differentiation in (B12) is licit. Now, by (B8) and (B9), one has $u_a \partial_T u_a = O(1/r^3)$ at large $r$, hence the limit on the r.h.s. of (B12) is zero. Thus:

$$\int_{\text{space}} u_{a,i} \, \partial_T^2 u_a \, dV = (d/dT) \int_{\text{space}} \partial_i u_a \partial_T u_a \, dV \,. \quad (B17)$$

Now we shall shew that

$$\int_{\omega_a} 4\pi G\rho \, (\partial^2 w_a/\partial x^i \partial T) \, dV = \int_{\text{space}} \partial_i u_a \partial_T u_a \, dV \,. \quad (B18)$$

Together with (B17), this will prove (A19). We substitute $-\Delta u_a$ for $4\pi G\rho$ and use Green's formula:

$$\int_{\omega_a} 4\pi G\rho \, (\partial^2 w_a/\partial x^i \partial T) \, dV = \lim_{r \to \infty} \int_{B_r} -\Delta u_a \, (\partial^2 w_a/\partial x^i \partial T) \, dV = \lim_{r \to \infty} \int_{B_r} -u_a \Delta(\partial^2 w_a/\partial x^i \partial T) \, dV +$$

$$+ \lim_{r \to \infty} \int_{S_r} [u_a(\partial/\partial \mathbf{n})(\partial^2 w_a/\partial x^i \partial T) - (\partial^2 w_a/\partial x^i \partial T)\partial u_a/\partial \mathbf{n}] dS \equiv \lim_1 + \lim_2 \quad (B19)$$

($B_r$ is the ball $|\mathbf{x} - \mathbf{a}| \leq r$ and $S_r$ the sphere $|\mathbf{x} - \mathbf{a}| = r$). We use the divergence theorem:

$$\lim_1 = \lim_{r \to \infty} \int_{B_r} -u_a(\partial^2 u_a/\partial x^i \partial T) \, dV = \lim_{r \to \infty} [\int_{S_r} -u_a(\partial_T u_a) \, n_i \, dS + \int_{B_r} u_{a,i} \, \partial_T u_a \, dV], \quad (B20)$$

and by (B8) and (B9) we get thus:

$$\lim_1 = \lim_{r \to \infty} \int_{B_r} u_{a,i} \, \partial_T u_a \, dV \equiv \int_{\text{space}} \partial_i u_a \partial_T u_a \, dV. \tag{B21}$$

From (B5), we find that $\partial^2 w_a /\partial x^i \partial T = O(1/r)$ and $\partial^3 w_a /\partial x^i \partial T \partial x^j = O(1/r^2)$. Since $\partial u_a/\partial x^i = O(1/r^2)$ and $u_a = O(1/r)$, we get $\lim_2 = 0$ and, with (B21), obtain (B18). This completes the proof.

## Appendix C. Calculation of $K^{ai}$ (very-well-separated spherical rigid bodies)

Let us calculate $K^{ai}$ [Eq. (A13)] under the two assumptions made at the end of §6.1. We begin with $K_1^{ai}$, using Eq. (6.13). Since the self-potential $u_a$ of body ($a$) is assumed spherical, we get inside ($a$):

$$k_{ij}(\mathbf{x}) = U h^{(a)}_{ij} [1 + O(\eta^2)] = [u_a(r) + U^{(a)}(\mathbf{x})] \, n_i \, n_j \, [1 + O(\eta^2)], \tag{C1}$$

where $\mathbf{n} = (n_i) \equiv (\mathbf{x} - \mathbf{a})/r$, $r \equiv |\mathbf{x} - \mathbf{a}|$. Since the pressure $p$ is also spherical, we have

$$\int_{\omega_a} u_a \, n_i \, n_j \, p_{,j} \, dV = \int_{\omega_a} u_a \, n_i \, n_j \, n_j \, p' \, dV = \int_{\omega_a} (u_a \, p')(r) \, n_i \, dV = 0, \tag{C2}$$

because $n_i$ has zero integral on each sphere $S_r$ (henceforth, the prime will denote derivative with respect to $r$, except for $K^{'ai}$). In the same way, we get

$$\int_{\omega_a} u_a \, n_i \, n_j \, \rho \, u_{a,j} \, dV = 0, \qquad \int_{\omega_a} p u_{a,i} \, dV = 0. \tag{C3}$$

To evaluate the contribution of the external potential $U^{(a)}(\mathbf{x})$, we make a first-order Taylor expansion of it at $\mathbf{x} = \mathbf{a}$ and use the following formula [5, Eq. (90.19)]:

$$\int n_i \, n_j \, d\omega = (4\pi/3) \, \delta_{ij} \qquad (d\omega \equiv \sin\theta \, d\theta \, d\varphi). \tag{C4}$$

We find thus:

$$K_1^{ai} = [(2/3)\varepsilon_a - \xi_a - \theta_a] \, U^{(a)}_{,i}(\mathbf{a}), \tag{C5}$$

where $\varepsilon_a$ is given by (5.12), and with

$$\xi_a \equiv -(1/3) \int_{\omega_a} \rho \, (\mathbf{x} - \mathbf{a}) \cdot \nabla u_a \, dV = -(4\pi/3) \int_{\omega_a} \rho \, u'_a \, r^3 \, dr, \tag{C6}$$

$$\theta_a \equiv -(1/3) \int_{\omega_a} (\mathbf{x} - \mathbf{a}) \cdot \nabla p \, dV = -(4\pi/3) \int_{\omega_a} p'_a \, r^3 \, dr. \tag{C7}$$

But, since Fock's Eqs. (73.16) and (73.19) [5] may be immediately rewritten as

$$\rho u_{a,i} + \Omega^{(a)}_{ik} \Omega^{(a)}_{jk} (x^j - a^j) = p_{,i}, \tag{C8}$$

we get by (5.10), independently of the sphericity assumption:

$$-3\xi_a + 2T_a = -3\theta_a. \tag{C9}$$

Therefore, (C5) may be written as

$$K_1^{ai} = [(2/3)(\varepsilon_a + T_a) - 2\xi_a]\, U^{(a)}{}_{,i}(\mathbf{a}). \tag{C10}$$

In the same way, using $(C3)_2$ and Taylor-expanding $p$ at $\mathbf{a}$, we get by (A5):

$$K_2^{ai} = \int_{\omega_a} p U^{(a)}{}_{,i}\, dV = (1/3)(\varepsilon_a - 2T_a)\, U^{(a)}{}_{,i}(\mathbf{a}). \tag{C11}$$

We also get (to the lowest order in $\eta$)

$$K_3^{ai} = -2[\int_{\omega_a} \rho\, u_a\, U^{(a)}{}_{,i}\, dV + \int_{\omega_a} \rho\, u_{a,i}\, U^{(a)}\, dV] = (2\xi_a - 4\varepsilon_a)\, U^{(a)}{}_{,i}(\mathbf{a}) \tag{C12}$$

and, using (5.4), (6.13) and (C4),

$$K'^{ai} = \int_{\omega_a} \rho\, (u_a + U^{(a)})\, n_i\, n_j\, (\dot{a}^j + \Omega^{(a)}{}_{jk}\, r n_k) dV = \dot{a}^i\, [2\varepsilon_a + M_a\, U^{(a)}(\mathbf{a})]/3. \tag{C13}$$

As to $K_4^{ai}$, we set $\mathbf{y} = \mathbf{x} - \mathbf{a}(T)$ and we may write using (C1):

$$K_4^{ai} = \int_{\omega_a} \rho u^j u^k [(u_a y^j y^k/r^2)_{,i} - (u_a y^i y^k/r^2)_{,j}] dV + \int_{\omega_a} \rho\, u^j u^k\, U^{(a)} [(y^j y^k/r^2)_{,i} - (y^i y^k/r^2)_{,j}]\, dV +$$

$$+ \int_{\omega_a} \rho\, u^j u^k [(U^{(a)}{}_{,i}\, y^j y^k - U^{(a)}{}_{,j}\, y^i y^k)/r^2] dV. \tag{C14}$$

Here, $u^j$ is given by (5.4), and $u_a = u_a(r)$. Moreover, $U^{(a)}$ may be replaced by its first-order Taylor expansion at $\mathbf{a}$, since we retain merely the lowest order in $\eta$ ; and we have

$$(y^j y^k/r^2)_{,i} - (y^i y^k/r^2)_{,j} = (y^j \delta_{ik} - y^i \delta_{jk})/r^2 = (n_j \delta_{ik} - n_i \delta_{jk})/r. \tag{C15}$$

Therefore, we still may write:

$$K_4^{ai} = \dot{a}^j \dot{a}^k\, (\int r\rho(r)\, u_a(r) dr)\, (\int (n_j \delta_{ik} - n_i \delta_{jk}) d\omega)$$

$$+ (\int r^2 \rho(r) u_a(r) dr)\, (\int (n_j \delta_{ik} - n_i \delta_{jk})(\Omega^{(a)}{}_{lj}\, n_l\, \dot{a}^k + \Omega^{(a)}{}_{lk}\, n_l\, \dot{a}^j) d\omega)$$

$$+ (\int r^3 \rho(r) u_a(r) dr)\, (\int (n_j \delta_{ik} - n_i \delta_{jk})(\Omega^{(a)}{}_{lj}\, n_l\, \Omega^{(a)}{}_{mk}\, n_m) d\omega)$$

$$+ U^{(a)}(\mathbf{a}) \iint r\rho(r)\, (n_j \delta_{ik} - n_i \delta_{jk})\, (\dot{a}^j + r\Omega^{(a)}{}_{mj}\, n_m)\, (\dot{a}^k + r\Omega^{(a)}{}_{pk}\, n_p)\, dr\, d\omega$$

$$+ U^{(a)}{}_{,l}(\mathbf{a}) \iint r^2 \rho(r)\, n_l\, (n_j \delta_{ik} - n_i \delta_{jk})\, (\dot{a}^j + r\Omega^{(a)}{}_{mj}\, n_m)\, (\dot{a}^k + r\Omega^{(a)}{}_{pk}\, n_p)\, dr\, d\omega$$

$$+ \iint r^2 \rho(r)\, [U^{(a)}{}_{,i}(\mathbf{a}) n_j n_k - U^{(a)}{}_{,j}(\mathbf{a}) n_i n_k]\, (\dot{a}^j + r\Omega^{(a)}{}_{mj}\, n_m)\, (\dot{a}^k + r\Omega^{(a)}{}_{pk}\, n_p)\, dr\, d\omega$$

$$= I_1 + I_2 + I_3 + I_4 + I_5 + I_6 \tag{C16}$$

(the $r$-integrals are from $r = 0$ to $r = r_a$). With the help of Eqs. (6.9), (C4), and [5, Eq.(90.20)]

$$\int n_i\, n_j\, n_k\, n_l\, \mathrm{d}\omega = (4\pi/15)\, (\delta_{ij}\delta_{kl} + \delta_{ik}\delta_{jl} + \delta_{il}\delta_{jk}), \tag{C17}$$

one finds in a straightforward manner that:

$$I_1 = I_3 = 0, \tag{C18}$$

$$I_2 = (4\pi/3)(\int r^2 \rho(r) u_a(r)\mathrm{d}r)[(\Omega^{(a)}{}_{lj}\, \dot{a}^i + \Omega^{(a)}{}_{li}\, \dot{a}^j)\delta_{jl} - 2\Omega^{(a)}{}_{lj}\, \dot{a}^j \delta_{il}] = 2\varepsilon_a\, \Omega^{(a)}{}_{ji}\, \dot{a}^j \equiv 2\varepsilon_a\, (\boldsymbol{\omega}_a \wedge \dot{\mathbf{a}})_i, \tag{C19}$$

$$I_4 = U^{(a)}(\mathbf{a})[\int r^2 \rho(r) u_a(r)\mathrm{d}r](4\pi/3)(\Omega^{(a)}{}_{ji}\, \dot{a}^j + \Omega^{(a)}{}_{jj}\, \dot{a}^i - \Omega^{(a)}{}_{ij}\, \dot{a}^j - \Omega^{(a)}{}_{ij}\, \dot{a}^j) = M_a U^{(a)}(\mathbf{a})(\boldsymbol{\omega}_a \wedge \dot{\mathbf{a}})_i, \tag{C20}$$

$$I_5 = U^{(a)}{}_{,l}(\mathbf{a})\, [(M_a/3)(\dot{a}^l \dot{a}^i - \delta_{il}\, \dot{a}^j \dot{a}^j) - (\gamma_a/5)(\Omega^{(a)}{}_{mj}\, \Omega^{(a)}{}_{mj}\delta_{il} + 2\Omega^{(a)}{}_{ij}\, \Omega^{(a)}{}_{lj})], \tag{C21}$$

$$I_6 = (M_a/3)[U^{(a)}{}_{,i}(\mathbf{a})\, \dot{a}^j \dot{a}^j - U^{(a)}{}_{,j}(\mathbf{a})\dot{a}^j \dot{a}^i] + (\gamma_a/5)[0 \times U^{(a)}{}_{,i}(\mathbf{a}) - 0 \times U^{(a)}{}_{,j}(\mathbf{a})] \tag{C22}$$

[$\gamma_a$ is given by Eq. (6.5)]. Summing this, one gets finally

$$K_4{}^{ai} = [M_a U^{(a)}(\mathbf{a}) + 2\varepsilon_a]\, \Omega^{(a)}{}_{ji}\, \dot{a}^j - (\gamma_a/5)[\Omega^{(a)}{}_{mj}\, \Omega^{(a)}{}_{mj}\, U^{(a)}{}_{,i}(\mathbf{a}) + 2\Omega^{(a)}{}_{ij}\, \Omega^{(a)}{}_{lj}\, U^{(a)}{}_{,l}(\mathbf{a})], \tag{C23}$$

or (using the formula of the double vector product)

$$\mathbf{K}_4{}^a = [M_a U^{(a)}(\mathbf{a}) + 2\varepsilon_a]\, \boldsymbol{\omega}_a \wedge \dot{\mathbf{a}} + (\gamma_a/5)\{2[\boldsymbol{\omega}_a \cdot \nabla U^{(a)}(\mathbf{a})]\, \boldsymbol{\omega}_a - 4\boldsymbol{\omega}_a{}^2\, \nabla U^{(a)}(\mathbf{a})\}. \tag{C24}$$